\documentclass[]{pasj02}

\jyear{2024}
\Received{}
\Accepted{}
 
\usepackage[switch,mathlines]{lineno}

\usepackage{ulem}

\begin{document} 

\title{X-ray Spectral and Timing Properties of the Black Hole Binary XTE J1859+226 and their Relation to Jets
  
}

\author{
Kazutaka \textsc{Yamaoka},\altaffilmark{1}\altemailmark\orcid{0000-0003-3841-0980} \email{yamaoka@isee.nagoya-u.ac.jp}
Toshihiro \textsc{Kawaguchi},\altaffilmark{2}\orcid{0000-0002-3866-9645}
Michael L. \textsc{McCollough},\altaffilmark{3}\orcid{0000-0002-8384-3374}
Ruben \textsc{Farinelli},\altaffilmark{4}\orcid{0000-0003-0536-4017}
 and 
Sergei \textsc{Trushkin}\altaffilmark{5,6}\orcid{0000-0002-7586-5856}
}
\altaffiltext{1}{Institute for Space-Earth Environmental Research (ISEE), Nagoya University, Furo-cho, Chikusa-ku, Nagoya 464-8601, Japan}
\altaffiltext{2}{Department of Economics, Management and Information Science, Onomichi City University, 1600-2, Hisayamada, Onomichi, Hiroshima 722-8506, Japan}
\altaffiltext{3}{Center for Astrophysics (CfA), Harvard University, B-240 60 Garden St. Cambridge, MA 02138, USA}
\altaffiltext{4}{INAF -- Osservatorio di Astrofisica e Scienza dello Spazio di Bologna, Via P. Gobetti 101, I-40129 Bologna, Italy}
\altaffiltext{5}{Special Astrophysical Observatory of the Russian Academy of
Sciences, Nizhnij Arkhyz, 369167, Karachayevo-Cherkessia, Russia}
\altaffiltext{6}{Kazan Federal University, Kremlyovskaya str., Kazan 420008,
Tatarstan, Russia}




\maketitle

\begin{abstract}

We compiled the X-ray and soft gamma-ray observations of the Galactic black hole binary XTE J1859+226 in the 1999--2000 outburst from RXTE, ASCA, BeppoSAX and CGRO. 
Throughout systematic spectral analysis using a two-component model consisting of a multi-temperature accretion disk plus a fraction of its flux convolved with an empirical Comptonized powerlaw component,  we found that the innermost radius ($r_{\rm in}$) and temperature ($T_{\rm in}$)  of the disk are very variable with time in the rising phase of soft X-ray flux where Type-A/B/C low-frequency quasi-periodic oscillations (QPOs) were found. After this phase, $r_{\rm in}$ remains constant at around 60 km assuming a distance of 8 kpc and an inclination angle of 67$^{\circ}$, and $T_{\rm in}$ smoothly decays with time. The constant $r_{\rm in}$ suggests a presence of the innermost stable circular orbit (ISCO), with $r_{\rm in}$ repeatedly moving closer and farther away from the ISCO in the rising phase.  Both disk parameters are remarkably correlated with independently analyzed timing properties such as QPO frequency and rms variability.  Type-A/B QPOs are seen only when $r_{\rm in}$ is close to the ISCO, while Type-C are seen when $r_{\rm in}$ is truncated and the frequency changes with a relation of $r^{-1.0}_{\rm in}$, supporting that Type-C QPOs occur at the inner edge of the truncated disk. Accurate determinations of the frequency--$r_{\rm in}$ relation for various objects should be a powerful tool to discriminate plausible Type-C QPO models. 
Furthermore, we suggest that jet ejection events may occur when $r_{\rm in}$ rapidly approaches to the ISCO, along with rapid changes of the disk flux, the rms variability and the hardness ratio. 
A rapid shrinkage of $r_{\rm in}$ down to the ISCO can be a useful index as a precursor of radio flares for triggering Target-of-Opportunity observations and would provide constraints on jet launching mechanisms.

\end{abstract}

\section{Introduction}


Black hole X-ray binaries (BHXBs) are close-binary systems consisting of a stellar mass black hole (BH) and a companion star. Matter from the companion star is accreted onto the BH due to angular momentum transfer and release gravitational potential energies at the vicinity of the BH. The matter is heated up to 10$^{7-9}$~K and mainly emits radiation in X-rays and soft gamma-rays (keV--MeV range).
Due to their proximity relative to the Earth (kpc scale) and relatively low masses (few to tens of M$_{\odot}$) the observed X-ray flux can get brighter than that of Crab Nebula and vary in much shorter time scale (ms$\sim$days) than that of Active Galactic Nuclei (AGN) with supermassive BH (weeks to $\sim$years). The BHXBs are classified into microquasars in analog to AGN in the extra-galactic, hence they are an ideal object for a study of an accretion flow onto a BH, BH physical properties (mass and spin) and its relation to relativistic jets (see \cite{review1, review2} for a review of BHXBs).

 Most of the BHXBs are classified as an X-ray transient except for a few persistent high mass systems like Cygnus X-1, LMC X-1 and LMC X-3. They tend to be discovered by X-ray all-sky monitor when the X-ray flux suddenly increases without any clear precursor event.  About 100 sources have been discovered so far and the list is summarized in several catalogs (\cite{watchdog,blackcat}).  The outburst of BHXBs typically lasts for several months to several years depending on sources, and evolves by changing spectral states: low/hard state (LHS), hard/soft intermediate state (HIMS/SIMS) or intermediate/very high state (IS/VHS), high/soft state (HSS) and quiescent states (QS) . 

At the beginning of the outburst which normally starts in the LHS,  which is dominated by a power-law with a photon index $\Gamma$ of 1.4--1.6  and an exponential cutoff at $\sim$100 keV in the X-ray spectrum.
This is considered to be a result of thermal Comptonization of soft photons from a cold accretion disk ($\sim$ 1 keV) by a geometrically-thick and optically-thin, hot accretion flow [Advection Dominated Accretion Flow (ADAF; \cite{ADAF,ADAF2}) or corona] with an electron temperature of $\sim$100 keV. As the X-ray flux increases, the source moves into the HSS via the HIMS and SIMS. The HSS is dominated by a blackbody-like thermal spectrum with a much cooler temperature of $\sim$1 keV with a softer power-law component with a photon index of 2.0--2.5 with no spectral break up to $\sim$ 100 keV or more. The thermal emission is perceived as 
radiation from a geometrically-thin and optically-thick accretion disk (standard accretion disk; \cite{standard}).
It is believed that the innermost radius ($r_{\rm in}$) of the accretion disk extends down to the innermost stable circular orbit (ISCO) in the HSS. However in the bright LHS, i.e., before the LHS-to-HSS  state transition, whether the $r_{\rm in}$ is truncated (e.g., \cite{ADAF2}) or it already reaches the ISCO (e.g. Lamppost model; \cite{lamppost}) is still debated.
After the HSS, the source goes back to the LHS via HIMS at a few \% of the Eddington limit, and then back into the QS.  The BHXB outburst has a hysteresis behaviour with a Q-shape in the hardness-intensity diagram (HIDs; \cite{hid}). More importantly, some BHXBs also exhibit large-scale relativistic jets from the BH, in addition to a compact jet seen in the LHS. It has been firmly established that the jets are ejected by passing through the LHS-to-HSS transition (so-called "jet line"; \cite{jet}), but no successful model so far explains physically how jets are launched.  
 
In addition to the X-ray spectral properties the X-ray timing properties such as power spectral densities (PSDs) and time lags are very crucial information on understanding accretion and jet mechanisms, given that they significantly change from state to state.  The X-ray light curves show the variability with a short time scale of msec in the LHS, while it is less variable in the HSS. Quasi-periodic oscillations (QPOs) are often observed as unique features in the HIMS/SIMS of BHXBs in the low frequency (LF) range (0.1--30 Hz) and also high frequency (HF) range ($>$50 Hz) of the PSDs [see \citet{review_qpo} for a QPO review], their origins are still debated. Several ideas have been proposed, e.g., accretion ejection instability (AEI; \cite{AEI}), Lense-Thirring (LT) 
precession (\cite{LT,LT2,Fragile2016,Bollimpalli2024}), 
accretion disk-oscillation (\cite{diskosc}). Based on the observed QPO properties (central frequency, $Q$ value, amplitude and noise), the LF-QPOs are classified into the three main types, Type-A, B and C (\cite{qpodef1,qpodef2,qpodef3}). Type-A and -B QPOs are seen in the SIMS, while Type-C QPOs are in the HIMS. Switches among different Types 
are very fast, occurring in a few seconds. Several authors suggested that Type-B QPOs were related to jet production mechanisms  (e.g., \cite{jet2}) and Type-C QPOs were related with the innermost radius of the accretion disk (\cite{maxi1659,Misra2020,Zhang2024}).

A new X-ray transient XTE J1859+226 was first discovered by the All-Sky Monitor (ASM) onboard Rossi X-ray Timing Explorer (RXTE) on October 9th, 1999 (MJD 51460; \cite{discovery}; hereafter we define this discovery date as origin), and was continuously monitored in X-rays by RXTE pointed instruments throughout the outburst lasting 9 months (\cite{rxte99}). The source was also observed by BeppoSAX (\cite{sax99}), Advanced Satellite for Cosmology and Astrophysics (ASCA), and Compton Gamma-Ray Observatory (CGRO; \cite{batse99}). The X-ray spectral and timing evolution were similar to typical outburst of BHXBs. 
In addition to several types of LF-QPOs (Type-A, B and C) in the range of 1--9 Hz (\cite{qpodef2}), the HF-QPO at roughly 187 and 150  Hz was discovered (\cite{hfqpo}) but later confirmed as low significance (\cite{summary_hfqpo}). The unique feature of the 1999--2000 outburst was that radio flares, probably due to jet ejections, occurred at least five times during the initial outburst phase (\cite{radio}). Hence, many authors have published spectral and timing results based on large RXTE data sets (\cite{1859_qpo,1859_qpo2,1859_sax,1859_crosscor,1859_spec}).

The source is located well above the Galactic plane [($l$,$b$)=(54.05,+8.61)], hence it has also been extensively studied in the optical and Ultra-Violet (UV) band due to its low extinction [$E(B-V)$=0.58$\pm$0.12; \cite{uv}]. Due to large uncertainties in measurement of the orbital period, the mass function was originally derived to be 7.4$\pm$1.1 M$_{\odot}$ (\cite{massfunc1}) and  further revised to 4.5$\pm$0.6 M$_{\odot}$ (\cite{massfunc2}), which gives a lower limit for the mass of the central object (a BH). Since the estimated value exceeds 3 M$_{\odot}$ (upper limit of the neutron star mass), this is the direct evidence for a presence of a BH in XTE J1859+226. Recently the BH mass ($M$) and inclination angle ($i$; with the face-on view corresponding to $i=0^{\circ}$) were accurately determined to be $M$=7.8$\pm$1.9 M$_{\odot}$ and $i$=66.6$\pm$4.3 degrees respectively by detailed optical spectroscopy of the H$_{\alpha}$ line (\cite{opt_mass}). Hereafter we assume these values for estimating the innermost radius ($r_{\rm in}$) in spectral parameters. On the other hand, the distance to this source ($D$) is still quite uncertain and different values have been proposed (7.6~kpc, \cite{uv}; 4.2~kpc, \cite{index_freq}; 11~kpc, \cite{optical}; 14~kpc, \cite{massfunc2}).  We follow $D$=8$\pm$3 kpc (\cite{uv2,watchdog}), which encompasses most of the past measurements.
 
In this paper, we compiled all the X-ray and soft gamma-ray observations of this source from RXTE, ASCA, BeppoSAX, and CGRO, and performed a systematic spectral study of XTE J1859+226 by modeling them with the disk blackbody convolved with the SIMPL Comptonization model (\cite{simpl}). Then we compared the spectral parameters with timing information, in section~3 and finally discuss its relation with plasma ejections (section~4).
  

\section{Observations and Data Analysis}
\subsection{RXTE}
A series of RXTE observations were carried out from October 11, 1999 (MJD 51462), i.e. two days after the discovery, to July 24, 2000 (MJD 51749) during quiescence. The observation ID starts with P40122, P40124, P40440, and P50401. The RXTE (\cite{rxte}) carries the two co-aligned instruments with narrow field of view of about 1 degree limited by collimators, the Proportional Counter Array (PCA; \cite{pca}) which covers the 2--100 keV range, the High Energy X-ray Timing Experiment (HEXTE; \cite{hexte}) which covers the 15--250 keV range, in addition to the ASM (\cite{asm}). The PCA consists of five identical proportional counter units referred to PCU 0-4, and the PCU 2
is the most active and the best calibrated detector among the five.  The HEXTE consists of two identical units (Cluster A and B). However, 
for the cluster B, spectral information from the analog-to-digital converter (ADC) was lost, leading to the reduction of the effective area by a factor of 3/4 compared to the cluster A. The Cluster A and B are operated  by changing its boresight directions to the source and a $+$/$-$1.5 degree-offset position for background estimation every 32 seconds. 

We used 131 observations from the RXTE data archive and selected the data with the standard criteria:
(1) the angle between the source and actual pointing direction is less than 0.02 degrees (OFFSET$<$0.02),
(2) elevation angle from the Earth is above 10 degrees (ELV$>$10), 
(3) the PCU 2 was active (PCU2\_ON==1) 
and (4) there was no breakdown during the observation period.
 
The PCA standard-2 data with 129 energy channels and 16-sec time resolution
were analyzed using the scripts {\tt pcaprepfile}, {\tt pcaextspect2} and {\tt pcaextlc2} in the HEASOFT 6.33.2. We collected the data from only the top layer of the PCU 2, and calculated the PCA background by either {\tt cmbright} or {\tt cmfaint} model based on recommended criteria of the source brightness (40 cts s$^{-1}$), corrected for deadtime for both source and background. The 0.5\% systematic error was added for each bin in the spectral analysis. The current response matrix gives the best-fit power-law value of $\Gamma$=2.11$\pm$0.01, and 2--10 keV flux of 2.24$\times$10$^{-8}$ erg cm$^{-2}$ s$^{-1}$ for Crab Nebula observation, which is almost consistent with canonical value (\cite{crab}). For the HEXTE data, exactly the same good time intervals (GTIs) as the PCA were applied. We used the standard archive data which has 64 energy channels with 16 second time-resolution.  The data were firstly separated into the source and offset ($+$/$-$) direction
 by the {\tt hxtback} script, and accumulated spectra from Cluster A (0123) and B (013)  by {\tt saextrct} and typical deadtime $\sim$40\% was corrected by {\tt hxtdead}. The backgrounds for the $+$/$-$ offset pointing were co-added.
 
\begin{figure*}
 \begin{center}
   \includegraphics[width=16cm]{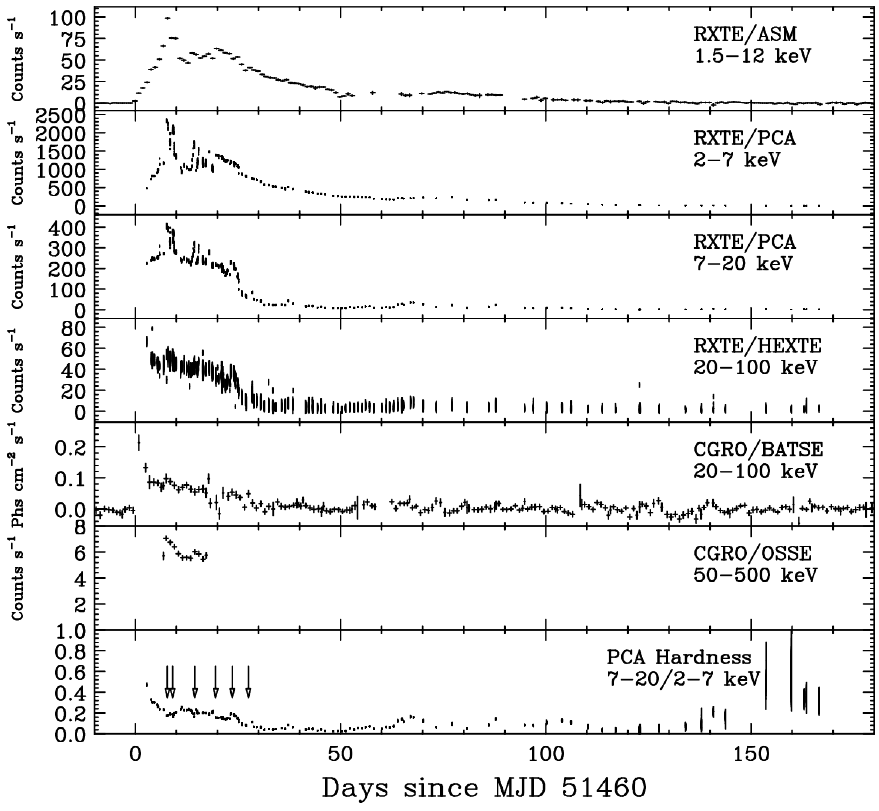}
 \end{center}
\caption{X-ray and soft gamma-ray light curves of XTE J1859+226 in the 1999-2000 outburst taken from several instruments: RXTE/ASM 1.5--12 keV, RXTE/PCA 2--7 keV and 7--20 keV, RXTE/HEXTE 20--100 keV, CGRO/BATSE 20--100 keV, CGRO/OSSE 50--500 keV,  and PCA hardness ratio between 2--7 and 7--20 keV from top to bottom. Six arrows in the bottom panel indicate radio flare peaks shown in \S3.4.}
\label{fig1:total_lc}
\end{figure*}

\begin{figure*}
 \begin{center}
 \includegraphics[width=8cm]{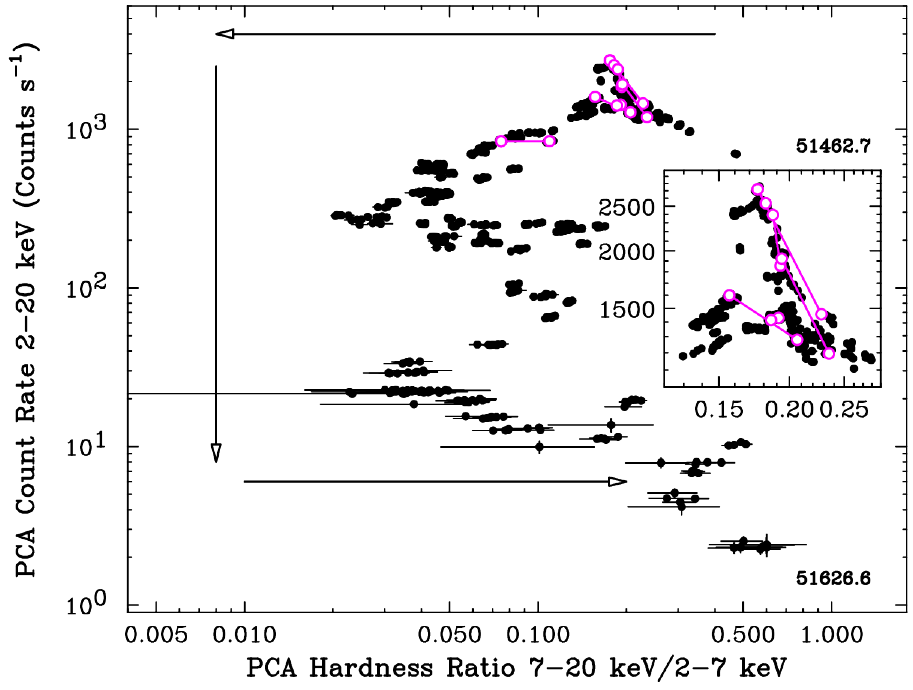} 
 \includegraphics[width=8cm]{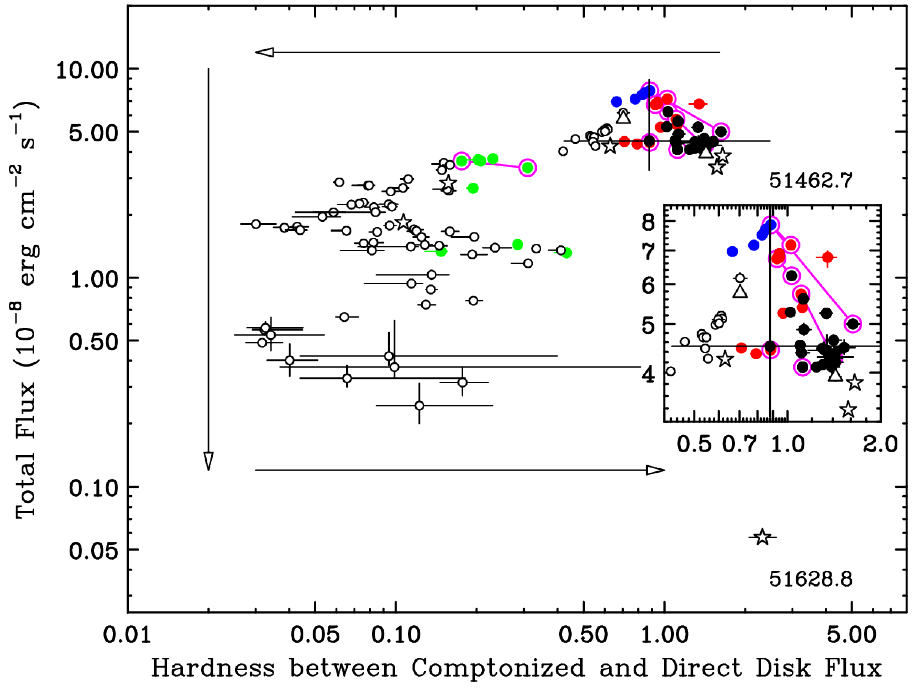}
 \end{center}
\caption{RXTE/PCA hardness-intensity diagram of XTE J1859+226 in the 1999-2000 outburst. The left figure is produced by the RXTE/PCA count rate, and the right is based on the spectral fitting results for several satellite data (RXTE shown by filled and  open circles with the same meaning as figures shown in \S3, ASCA+RXTE by open triangles, and BeppoSAX by open stars: see \S3.1 for definitions of hardness ratio and Comptonized/Direct Disk flux. Anti-clockwise arrows indicate the time evolution of the outburst in this plot. Numbers indicate the observation dates in MJD. Pairs of the purple circles connected by a line indicate 
the two epochs sandwiching the timing of each jet event 
(\S~4.3). Inserts show the expanded views around the jet ejection events.}
\label{fig2:hid}
\end{figure*}

Thus derived, total PCA exposure was 338.6 ksec, meaning that the averaged exposure was 2.59 ksec per one observation. The RXTE/PCA and HEXTE light curves, and the PCA HID  during the 1999--2000 outburst are shown together with CGRO/BATSE and OSSE data in figures \ref{fig1:total_lc} and \ref{fig2:hid}. 

The PCA count rates at $\leqq 20$~keV reach their peaks on MJD~51467, while HEXTE and CGRO/BATSE count rates at $\geqq 20$~keV have peaks on MJD~51460-51461 prior to the PCA. Hereafter, terms upon the time evolution, such as "peak epoch" and "rising phase", refer to the PCA count rates. Time variation of the fluxes of spectral components, rather than the count rates, will be investigated later (see section~3.2).

\subsection{ASCA}
The ASCA Target-of-Opportunity (ToO) observations were carried out on October 23, 1999 (MJD 51474). The ASCA (\cite{ASCA}) carries four X-ray Telescopes (XRTs; \cite{xrt}), and has following two types of scientific instrument with imaging capability on each focal plane of the XRT. The Gas Imaging Spectrometer (GIS: \cite{gis}) consists of two identical xenon gas scintillation proportional counters (GIS2 and GIS3), and the Solid-state Imaging Spectrometer (SIS; \cite{sis}) consists of two identical X-ray Charged-Couple-Device (CCD; SIS0 and SIS1). Both instruments are sensitive to the energy ranges at 0.5(SIS)/0.7(GIS) keV--10 keV, and complementary in terms of their energy resolution and field of view.
At the beginning of the ASCA mission, the SIS had a much better energy resolution  ($\Delta E \sim$140 eV at 6 keV) than that of GIS ($\Delta E \sim$480 eV at 6 keV), but due to the radiation damage to the CCD chips, the SIS energy resolution had been degraded to $\sim$350 eV before this observation. We have checked the heavily pile-uped SIS data following the method described in \citet{sisana}, but there was no clear emission/absorption-like structure in the 5--10 keV band. Hence, we determined to focus on the GIS data, which provides us more reliable continuum information, in this paper.

\begin{figure}
 \begin{center}
\includegraphics[width=8cm]{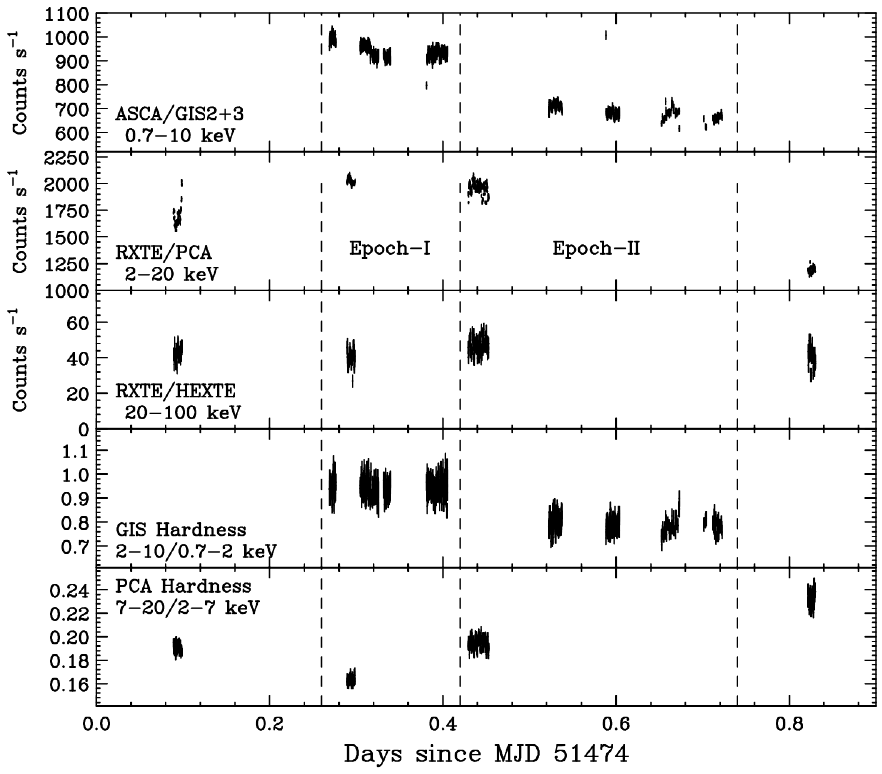} 
 \end{center}
\caption{ASCA/GIS, RXTE/PCA, and HEXTE light curves of XTE J1859+226 on September 26 (MJD 51474). 0.7--10 keV from ASCA/GIS(GIS2+3), 2--20 keV from RXTE/PCA(PCU2), 20--100 keV from RXTE/HEXTE(Cluster A), and the GIS hardness ratio between 0.7--2 keV and 2--10 keV, and PCA hardness ratio between 2--7 keV and 7--20 keV from top to lower. Dashed lines indicate two epochs we performed spectral analysis simultaneously with ASCA and RXTE.}\label{fig3:ascaxte_lc}
\end{figure}

 We utilized only the screened GIS event file with high bit-rate, because the medium bit-rate data were severely suffered from telemetry saturation.
 The GIS data were extracted within 6 arc-minutes of the image center by {\tt xselect}, and background was not subtracted due to the brightness of this source. The deadtime was corrected by {\tt deadtime}. The deadtime-corrected exposure was 2.51 and 2.13 ksec for the GIS2 and 3 respectively. For the spectral analysis, we co-added GIS2 and GIS3 data, and added a 2\% systematic error for each spectral bin.
The averaged GIS2+3 flux in the 2--10 keV range was 1.71$\times$10$^{-8}$ erg cm$^{-2}$ s$^{-1}$, corresponding to about 0.8 Crab. The GIS2+3 co-added light curves in the 0.7--10 keV range and the hardness ratio between 0.7--2 keV and 2--10 keV are shown in figure \ref{fig3:ascaxte_lc}, together with four quasi-simultaneous RXTE observations (obsID: 40124-01-15-02, 15-03, 24-00, and 25-00). The GIS hardness ratio significantly changed after MJD51474.41, which means that the spectral shape changed. Hence we separated the entire observation into two epochs refereed to the Epoch I and II shown in figure \ref{fig3:ascaxte_lc}. The Epoch-I and II include one RXTE observation 40124-01-15-03 and 40124-24-00 respectively, and spectral data were analyzed simultaneously within each Epoch. 

\subsection{BeppoSAX}
BeppoSAX observations were carried out six times as a ToO (referred as TOO1-6 in time order) during the time span from October 15, 1999 (MJD 51466) to March 26, 2000 (MJD 51629). Table \ref{tab1:obslog} shows the list of BeppoSAX observations together with ASCA and RXTE. The Beppo-SAX (\cite{sax}) has four narrow field instruments (NFIs) with very broadband energy coverage and low background: Low Energy Concentrator (LECS; 0.1--10 keV), Medium Energy Concentrator (MECS; 1.3--10 keV), High Pressure Gas Scintillation Proportional Counter (HPGSPC; 4--100 keV), and Phoswitch Detector System (PDS; 15--300 keV), hence BeppoSAX is very suitable for spectral study of the BHXBs which emits in the wide range from  soft X-ray to soft gamma-rays. In this paper, exactly the same data and products as published in \citet{1859_sax} were used  (see \cite{1859_sax} for details of observations and data extractions). Since the BeppoSAX has a sensitivity in lower energy below $\sim$3 keV, which is the low energy limit of the RXTE/PCA in spectral fitting, a more accurate measurement of the disk component is possible. RXTE results should be compared with the BeppoSAX results.

\subsection{CGRO/OSSE}
The Oriented Scintillation Spectroscopy Experiment (OSSE; \cite{osse}) onboard the CGRO observed XTE J1859+226 for almost twelve days from 16:10 (UT) on October 15, 1999 (MJD 51466) to 13:49 (UT) on October 26, 1999 (MJD 51477), which is a typical CGRO exposure for an individual source. The OSSE has four identical NaI(Tl)/CsI(Na) phoswich scintillators (OSSE 1234), which are sensitive to soft gamma-rays in the range of 50~keV--10~MeV. The four OSSE detectors can work independently and the primary and secondary target was set at XTE J1859+226 and Sun,  respectively. The deadtime-corrected exposure was 680.9 ksec. We used the high level data products; i.e., the background-subtracted,  deadtime-corrected spectrum and the response FITS file applicable to the XSPEC (\cite{xspec}) analysis, which are publicly available from the NASA/HEASARC. As can be seen from figure \ref{fig1:total_lc}, the OSSE observation was carried out when the hard X-rays above 20 keV was strong. The averaged OSSE flux was 1.795$\times$10$^{-9}$ erg cm$^{-2}$ s$^{-1}$ in the 50--500 keV range, corresponding to about 110 mCrab\footnote{The OSSE observation of Crab nebula/pulsar on 2000 Apr-May gives an energy flux of 1.688$\times$10$^{-8}$ erg cm$^{-2}$ s$^{-1}$ in the same energy range.}. 

\begin{table*}
\caption{Summary of observations used in this paper.}\label{tab1:obslog}
{\footnotesize
\begin{tabular}{llccccc}\\\hline
      Satellite & Instruments & ObsId & Start Time UT (MJD) & Stop Time UT (MJD) & Exposure\footnotemark[$*$] & Flux\footnotemark[$\dag$] \\    \hline
RXTE & PCA       & 40122,40124 & 1999-10-11 18:21:52 (51462.77) & 2000-07-24 05:14:40 (51749.22)  & 338.6 &  \\
       & HEXTE Cl. A\&B     & 40440,50401 &            &                               & 114.1 &  \\\hline
      ASCA & GIS2,GIS3    & 1570900 & 1999-10-23 06:01:19 (51474.25) & 1999-10-23 17:20:35 (51474.72) & 4.65 & 17.1 \\\hline
      BeppoSAX & MECS,HPGSPC,PDS & 21027001 & 1999-10-15 02:16:54 (51466.10) & 1999-10-15 09:43:22 (51466.41) &  14.1/7.1  & 12.4\\
            &   MECS,HPGSPC,PDS     &210270011  & 1999-10-22 00:45:04 (51473.03) & 1999-10-22 09:51:40 (51473.41) & 16.1/7.7   & 10.9\\
             &  LECS,MECS,HPGSPC,PDS &210270012 & 1999-10-28 19:46:36 (51479.82) & 1999-10-28 22:47:44 (51479.95) & 4.5/1.8  & 14.8\\
              & LECS,MECS,HPGSPC,PDS &210270013 & 1999-11-07 17:05:40 (51489.71) & 1999-11-08 12:55:12 (51490.54) & 29.1/13.6  & 8.04 \\
              & LECS,MECS,HPGSPC.PDS &210270014 & 1999-11-19 04:10:31 (51501.17) & 1999-11-20 00:47:45 (51502.03) & 30.1/13.6  & 4.58 \\
              & LECS,MECS,HPGSPC,PDS &210270015 & 2000-03-25 04:23:58 (51628.18) & 2000-03-26 12:33:13 (51629.52) & 48.8/21.4 & 0.0757\\\hline
      CGRO & OSSE & 99288001 & 1999-10-15 16:10:30 (51466.67) & 1999-10-26 13:49:21 (51477.58) & 680.9 & 1.80 \\\hline
\end{tabular}
\begin{tabnote}
\footnotemark[$*$] In the unit of ksec. PCA and HEXTE Cluster A for RXTE, GIS2+3 for ASCA, and MECS and PDS for BeppoSAX.\\
\footnotemark[$\dag$] In the unit of $10^{-9}$~erg cm$^{-2}$ s$^{-1}$. Measured in 2--10 keV for ASCA and BeppoSAX, 50--500 keV for OSSE.\\
\end{tabnote}
}
\end{table*}

\section{Analyses and Results}

\subsection{Spectral Fitting}

To perform systematic spectral analysis throughout this paper, we chose the multi-color disk-blackbody (MCD) model (\cite{mcd1,mcd2}; {\tt diskbb} in XSPEC) convolved with the SIMPL Comptonization (\cite{simpl}) as the continuum. This convolution model can provide more accurate {\tt diskbb} parameters (innermost temperature $T_{\rm in}$ and innermost radius $r_{\rm in}$) in comparison with the conventional {\tt diskbb} plus independent Comptonization model. Such models ignore the contribution of disk photons to the Comptonized emission like power-law/cut-off power-law, or utilize seed photons inputted from different blackbody/disk-blackbody continuum like thermal Comptonization ({\tt nthcomp}; \cite{nthcomp}) and non-thermal Comptonization model including bulk-motion in relativistic flow ({\tt bmc}; \cite{bmc}). This independent treatment, just as an approximation for the spectral shape of a hard power-law component dominated above 10--20 keV, could result in underestimation of disk normalization  unless it is corrected (e.g., \cite{1550_spec2,simpl}). Spectral models used in previous works on this source are as follows; {\tt diskbb+powerlaw} (\cite{BHspec_sysstudy,1859_qpo2}), {\tt diskbb+bmc} (\cite{1859_sax}) and {\tt diskbb+nthcomp+rellxilCp} (where {\tt rellxilCp} is the relativistic reflection model; \cite{1859_xrs}). The estimation of disk fluxes and $r_{\rm in}$s using these models are probably underestimated.
 
 The continuum was further modified with Galactic absorption {\tt TBabs} model with solar abundances (\cite{abund_wilms}) and cross-section (\cite{cross_verner}),  and the {\tt smedge} model (\cite{smedge}) as an approximation of broad reflection-like structures above 7 keV. At the initial outburst phase (Day 2.7--6.0 and 11.0--13.0 in figure~\ref{fig1:total_lc}), we found that a high energy cutoff is required in 13 observations ($\chi^2/{\rm dof} = 1.85$). Hence we used the {\tt simplcutx} model \footnote{https://jfsteiner.com/?p=71}which is an extension version of the {\tt simpl} model and has an exponential cutoff of $\exp(-E/E_{\rm f})$ where $E_{\rm f}$ is the e-folding energy. 
 Thus, we applied the model {\tt constant$*$TBabs$*$smedge$*$simplcutx$\otimes$diskbb} (hereafter model A; c.f., model B in section~3.2)  which have eight free parameters in the model fitting.
 The diskbb model has two free parameters: normalization ($N_{\rm dbb}$) and $T_{\rm in}$. The normalization is related to $r_{\rm in}$ by $N_{\rm dbb}=r_{\rm in}^2 \cos{i}/D_{10}^2$ where $i$ is the inclination angle from the face-on view
 and $D_{10}$ is the source distance in unit of 10 kpc. The {\tt simplcutx} model has three free parameters of photon index ($\Gamma$),  Compton fraction ($f_{\rm sc}$) and e-folding energy ($E_{\rm f}$). In this model, a fraction ($f_{\rm sc}$) of photons from the {\tt diskbb} component are scattered by the Comptonizing medium and produces a power-law hard tail, while other remained fractions (1--$f_{\rm sc}$) are escaped from the medium and the disk component is directly seen by observers. Only up-scattering is considered and reflection of the scattering medium is not taken into account (Refl\_par=0). After MJD 51572, $E_{\rm f}$ was fixed at 10$^{6}$ keV (i.e., no cutoff) because they are not well constrained. 
 
 The hydrogen column density ($N_{\rm H}$) in the {\tt TBabs} model was fixed at the Galactic value of 3.55$\times$10$^{21}$ cm$^{-2}$ (\cite{nh}), and in the {\tt smedge} model, the edge energy ($E_{\rm edge}$; constrained above 7.11 keV due to cold iron-K edge) and optical depth ($\tau$) were left free with fixed default values for the width (10 keV) and the index ($-$2.67). We also changed $N_{\rm H}$ in the range of (2.3--3.6)$\times 10^{21}$ cm$^{-2}$ measured with ASCA and BeppoSAX shown in table \ref{tab2:fit_ascaxtesax}, and found that there were small differences of 0.5\% in $T_{\rm in}$ and 2\% in $r_{\rm in}$, respectively. The relative normalization among instruments ({\tt constant}) was fixed at 1 for the PCA, while it was left free for the HEXTE cluster A and B in the {\tt constant} model. As mentioned above, there are two spectral components: direct disk component and Comptonized component. The bolometric flux was integrated over 1 eV--1 MeV and estimated by using the {\tt cflux} model  based on the PCA normalization. We also estimated two disk fluxes: direct disk flux and total disk flux including contributions to Compton scattering. The disk flux was corrected by 1/(2$\cos{i}$) assuming the thin disk geometry. The spectral hardness ratio (SHR) estimated from spectral fits is defined as the ratio of the Comptonized flux relative to the direct disk flux.
 
 In order to fit the spectra and obtain the best-fit parameters,  we used the minimum-$\chi^2$ technique in the XSPEC version 12.14.0h, and quoted statistical errors at 68\% confidence level ($\Delta\chi^2$=1.0).  The averaged reduced $\chi^2$ was 1.06, indicating that the fitting quality is reasonable. 
  
\begin{figure*}
 \begin{center}
\includegraphics[width=16cm]{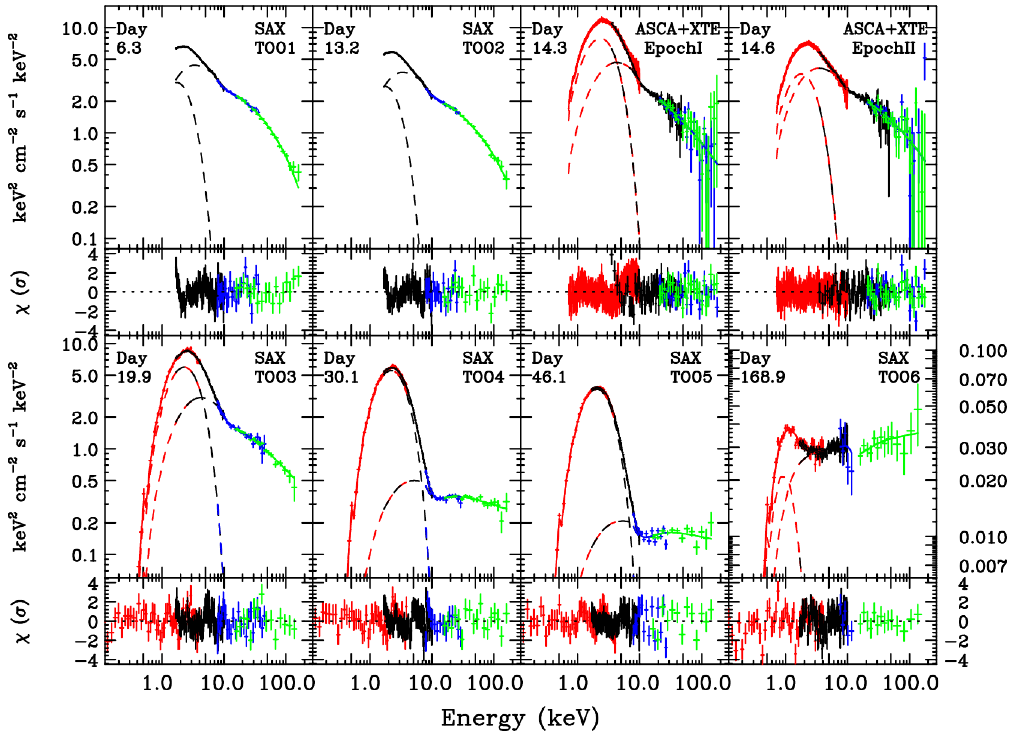} 
 \end{center}
\caption{Spectral fitting of simultaneous ASCA and RXTE data during Epoch-I and II, and BeppoSAX data during TOO1 to TOO6. Each upper panel shows the $\nu F_\nu$ spectra with the best-fit model A. The normalization factor among instruments is corrected. Models of direct-disk and Comptonized components are shown by dashed line. Each Lower panel shows residuals from the best-fit model. The figures are arranged in time order from top left to bottom right so that one can understand the time evolution of the energy spectrum. The different Y-axis scale is used in four top panels, three bottom left panels, one bottom-right panel for clarity of spectral shape.  }\label{fig4:ascaxtesax_fit}
\end{figure*}

We also show ASCA and BeppoSAX fitting results in figure \ref{fig4:ascaxtesax_fit}.  The ASCA/GIS2+3 and RXTE/PCA (PCU 2), HEXTE (Cluster A and B) data are simultaneously fitted with the model A for the Epoch I and II. The BeppoSAX data are also fitted with the same model. The MECS normalization was fixed at 1, and others (LECS, HPGSPC, and PDS) were set free. 
The best-fit parameters are summarized in table \ref{tab2:fit_ascaxtesax}. 
\begin{table*}
{\footnotesize
\caption{Best-fit parameters in simultaneous ASCA and RXTE (Epoch-I,II) and BeppoSAX observations (TOO1-6).}\label{tab2:fit_ascaxtesax}
\begin{tabular}{llcccc}\\\hline
Model & Parameters & TOO1 & TOO2 & Epoch-I & Epoch-II  \\\hline
TBabs   & $N_{\rm H}$(10$^{22}$cm$^{-2}$)	& 0.355 (fixed)	& 0.355 (fixed)	& 0.263$^{+0.007}_{-0.006}$	& 0.228$^{+0.008}_{-0.008}$	\\
diskbb & $T_{\rm in}$(keV)	& 0.624$^{+0.006}_{-0.006}$ 	& 0.636$^{+0.006}_{-0.006}$ 	& 0.924$^{+0.008}_{-0.008}$ 	& 0.681$^{+0.008}_{-0.008}$ 	\\
       & $r_{\rm in}$(km)	& 99.3$^{+2.1}_{-1.9}$ 	& 90.2$^{+1.8}_{-1.7}$ 	& 61.0$^{+1.0}_{-1.0}$ 	& 85.2$^{+2.1}_{-2.0}$ 	\\
simplcutx    & $\Gamma$	& 2.37$^{+0.01}_{-0.01}$ 	& 2.40$^{+0.01}_{-0.01}$ 	& 2.57$^{+0.04}_{-0.04}$ 	& 2.42$^{+0.03}_{-0.03}$ 	\\
             & $E_{\rm f}$(keV)	& 113.8$^{+6.3}_{-5.5}$ 	& 167.0$^{+13.1}_{-11.7}$ 	& 583.0$^{+19521.1}_{-294.6}$ 	& 343.0$^{+411.5}_{-125.2}$ 	\\
             & $f_{\rm SC}$ 	& 0.428$^{+0.008}_{-0.008}$	& 0.416$^{+0.008}_{-0.008}$	& 0.261$^{+0.011}_{-0.010}$	& 0.387$^{+0.013}_{-0.013}$	\\
smedge & $E_{\rm Edge}$(keV)	& 7.11$^{+0.02}_{-0.00}$ 	& 7.11$^{+0.02}_{-0.00}$ 	& 7.74$^{+0.11}_{-0.11}$ 	& 7.19$^{+0.13}_{-7.19}$ 	\\
         & $\tau$	& 1.40$^{+0.07}_{-0.07}$ 	& 1.43$^{+0.07}_{-0.07}$ 	& 1.39$^{+0.10}_{-0.10}$ 	& 1.17$^{+0.09}_{-0.09}$ 	\\
constant & factor GIS/LECS	& --	& --	& 0.892$^{+0.018}_{-0.018}$ 	& 0.934$^{+0.020}_{-0.020}$ 	\\
         & factor HEXTE Cl.A/HPGSPC	& 1.164$^{+0.007}_{-0.007}$ 	& 1.162$^{+0.008}_{-0.007}$ 	& 0.938$^{+0.021}_{-0.021}$ 	& 0.940$^{+0.023}_{-0.023}$ 	\\
         & factor HEXTE Cl.B/PDS	& 0.976$^{+0.009}_{-0.004}$ 	& 0.988$^{+0.010}_{-0.009}$ 	& 0.977$^{+0.004}_{-0.004}$ 	& 1.095$^{+0.004}_{-0.004}$ 	\\
$\chi^2$/d.o.f. & 	& 101.8/110	& 98.4/106	& 320.3/344	& 241.6/344	\\\hline
Model & Parameters & TOO3 & TOO4 & TOO5 & TOO6  \\\hline
TBabs   & $N_{\rm H}$(10$^{22}$cm$^{-2}$)	& 0.353$^{+0.007}_{-0.007}$	& 0.353$^{+0.005}_{-0.005}$	& 0.337$^{+0.004}_{-0.004}$	& 0.237$^{+0.022}_{-0.020}$	\\
diskbb & $T_{\rm in}$(keV)	& 0.882$^{+0.007}_{-0.008}$ 	& 0.842$^{+0.002}_{-0.002}$ 	& 0.770$^{+0.003}_{-0.003}$ 	& 0.255$^{+0.010}_{-0.010}$ 	\\
       & $r_{\rm in}$(km)	& 57.7$^{+1.0}_{-0.9}$ 	& 55.0$^{+0.4}_{-0.4}$ 	& 53.6$^{+0.5}_{-0.5}$ 	& 56.2$^{+7.4}_{-6.3}$ 	\\
simplcutx    & $\Gamma$	& 2.53$^{+0.03}_{-0.04}$ 	& 2.19$^{+0.02}_{-0.02}$ 	& 2.13$^{+0.03}_{-0.03}$ 	& 1.94$^{+0.02}_{-0.02}$ 	\\
             & $E_{\rm f}$(keV)	& $>$519.2	& 10$^6$ (fixed)	& 10$^6$ (fixed)	& 10$^6$ (fixed)	\\
             & $f_{\rm SC}$ 	& 0.230$^{+0.010}_{-0.009}$	& 0.044$^{+0.001}_{-0.001}$	& 0.027$^{+0.001}_{-0.001}$	& 0.242$^{+0.018}_{-0.018}$	\\
smedge & $E_{\rm Edge}$(keV)	& 7.96$^{+0.17}_{-0.18}$ 	& 7.66$^{+0.09}_{-0.08}$ 	& 7.62$^{+0.12}_{-0.11}$ 	& 7.11 (fixed) 	\\
         & $\tau$	& 1.53$^{+0.14}_{-0.13}$ 	& 2.43$^{+0.17}_{-0.17}$ 	& 2.66$^{+0.27}_{-0.27}$ 	& --	\\
constant & factor GIS/LECS	& 0.975839	& 1.133$^{+0.006}_{-0.006}$ 	& 1.027$^{+0.006}_{-0.006}$ 	& 0.740$^{+0.011}_{-0.011}$ 	\\
         & factor HEXTE Cl.A/HPGSPC	& 1.107$^{+0.012}_{-0.012}$ 	& 1.119$^{+0.012}_{-0.012}$ 	& 1.212$^{+0.022}_{-0.022}$ 	& 1.662$^{+0.122}_{-0.121}$ 	\\
         & factor HEXTE Cl.B/PDS	& 0.923$^{+0.016}_{-0.015}$ 	& 0.857$^{+0.016}_{-0.016}$ 	& 0.805$^{+0.029}_{-0.028}$ 	& 1.204$^{+0.059}_{-0.057}$ 	\\
$\chi^2$/d.o.f. & 	& 181.2/158	& 160.6/152	& 144.8/147	& 180.4/139	\\\hline
\end{tabular}
\begin{tabnote}
Errors are quoted at statistical 68\%.\\
\end{tabnote}
}
\end{table*}

\subsection{Time Evolution of Spectral Properties}

\begin{figure*}
\begin{center}
\includegraphics[width=16cm]{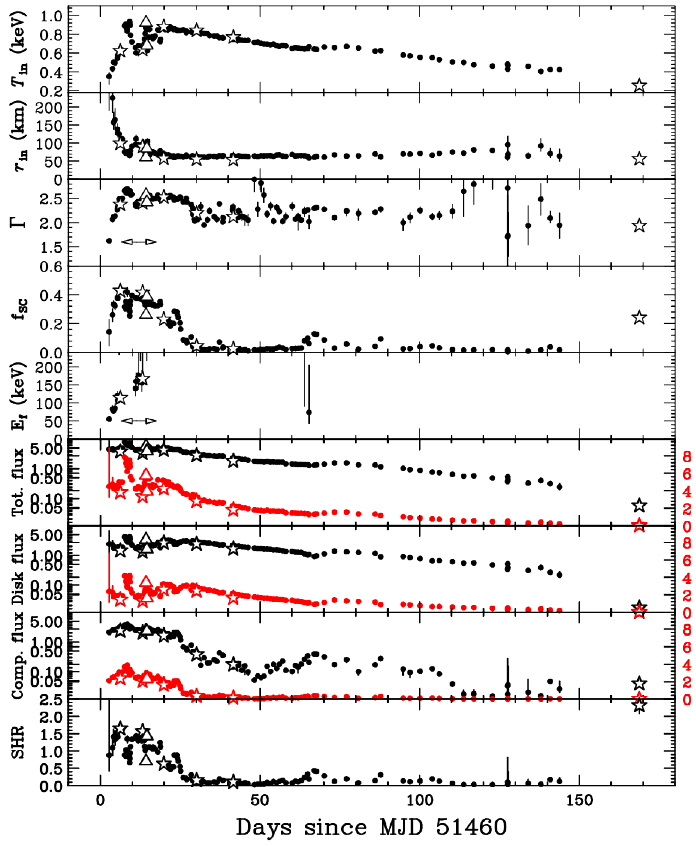} 
 \end{center}
\caption{Time evolution of spectral parameters. Top panel: Innermost temperature ($T_{\rm in}$) and innermost radius ($r_{\rm in}$) in {\tt diskbb}, photon index ($\Gamma$) and Compton scattering fraction ($f_{\rm sc}$), and e-folding energy ($E_{\rm f}$) in {\tt simplcutx}, total flux, disk direct flux and Comptonized flux, and spectral hardness ratio (SHR) between Comptonized flux to disk direct flux are shown from top to bottom. Double-sided arrow indicates the epochs when the CGRO/OSSE observations are carried out. The fluxes are shown in unit of 10$^{-8}$ erg cm$^{-2}$ s$^{-1}$ in both linear (red with the right axis) and logarithmic scale (black with the left axis). The results from ASCA+RXTE and BeppoSAX observations are indicated by open triangles and open stars, respectively. }\label{fig5:timevar_specpar}
\end{figure*}

Figure \ref{fig5:timevar_specpar} shows time evolution of spectral parameters obtained from the model fitting. The ASCA and BeppoSAX results described above are included in this figure. During the initial rising phase until MJD 51480 when $f_{\rm SC}$ is higher than 0.2, the $T_{\rm in}$ gradually rise from 0.36 keV up to 0.94 keV at maximum on MJD 51469.24, and then repeated up and down in the range of 0.6--0.9 keV several times for about 10 days. In association with the $T_{\rm in}$ time variation, $r_{\rm in}$ also changes from a large radius of 300 km to about 60 km. The Comptonized flux is more stable than the direct disk flux over $\sim 25$~days from the beginning of the observations, and then starts to decline. The behaviour of the SHR will be discussed later (section~4.3) in terms of the "jet line". 
  
After MJD 51480, 
$T_{\rm in}$ smoothly decayed from $\sim$0.9~keV down to $\sim$0.3~keV, and $r_{\rm in}$ was kept remarkably constant at about 60 km during the decay phase. The constant $r_{\rm in}$ during the outburst is one of unique features in various BHXBs. This fact is broadly interpreted to be due to a presence of the innermost stable circular orbit (ISCO: e.g., \cite{const_rin,const_rin2}). 

$\Gamma$ ranges from 1.6 at the beginning at $\sim$MJD~51462 to a steep value of 2.3--2.7 during 
 MJD~51465-51480. The e-folding energy $E_{\rm f}$ also gradually increases with time from 55 keV to 200 keV or more, and was not well constrained in MJD 51475--51524. 
This suggests either that electron temperature becomes higher than the instrumental energy coverage of RXTE/BeppoSAX or that the electron distribution in the corona/ADAF is in the transition process from purely thermal in the LHS to non-thermal in the HSS, e.g., hybrid thermal plus non-thermal distribution (\cite{hybrid_plasma}).

\begin{figure}
 \begin{center}
\includegraphics[width=8cm]{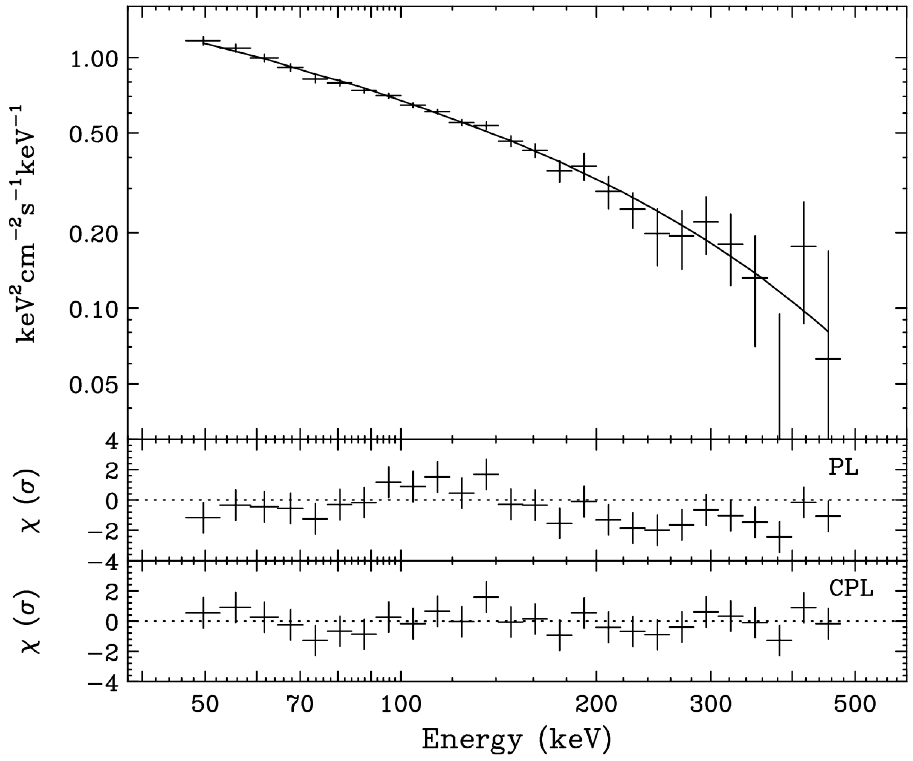} 
 \end{center}
\caption{Model fitting of the CGRO/OSSE 50--500 keV spectrum taken on MJD 51466--51477 (see figure \ref{fig1:total_lc}). Upper panel: $\nu F_{\nu}$ spectrum together with the best-fit cut-off power-law (CPL) model shown by solid line. Two lower panels: residuals from the best-fit power-law (PL) and CPL. }\label{fig6:osse}
\end{figure}

 Then, we also checked the OSSE data taken during the initial phase. The OSSE spectrum in the 50--500 keV range can be fitted by a single power-law (PL) giving a photon index of 2.87$\pm$0.03 ($\chi^2$/dof = 36.4/24).  We found if we used a cutoff power-law (CPL) model instead, the fit is greatly improved ($\chi^2$/dof=12.8/23). The best-fit parameters are a power-law photon index of 2.46$\pm$0.10 and the folding energy of 249$^{+74}_{-49}$ keV. Figure \ref{fig6:osse} shows the OSSE $\nu F_{\nu}$ spectrum with the best-fit model. The OSSE-measured photon index is consistent with the results obtained from RXTE observations. We also investigated time variability during the OSSE observation, and found that the photon indices are consistent with RXTE observations, but did not find any significant change of the folding energy. 

The peak flux in the 1999--2000 outburst was 7.86$\times$10$^{-8}$ erg cm$^{-2}$ s$^{-1}$ on MJD 51467.6, corresponding to 6.02$\times$10$^{38}$ erg s$^{-1}$ and 61.7\% of the Eddington luminosity $L_{\rm E}$ (=1.25$\times$10$^{38} M/M_{\odot}$ erg s$^{-1}$) for assumed distance of 8 kpc and the BH mass of 7.8 $M_{\odot}$. 

Results from RXTE data alone were also confirmed by simultaneous ASCA and RXTE observations, and BeppoSAX observations. The results for these additional data are consistent with RXTE observations except for 
 the $r_{\rm in}$ in BeppoSAX data which gives systematically slightly smaller value than RXTE.  We checked the Crab Nebula data and found that the BeppoSAX MECS flux is smaller by $\sim$20\% than the PCA. This value is reasonably consistent with the flux difference between RXTE and BeppoSAX.

 The last data point obtained with BeppoSAX (TOO6), when the bolometric flux is close to a minimum ($\sim$5$\times$10$^{-10}$ erg cm$^{-2}$ s$^{-1}$), shows $T_{\rm in}$, $r_{\rm in}$, $\Gamma$ and the disk flux at their extrapolations of other epochs (i.e., before the last data point). In this sense, the BeppoSAX TOO6 data seems to belong to the HSS. However, the HID compiled through various satellites indicates that the data are located at lower-right corner, which is comparable SHR of the SIMS/HIMS  at the upper-right corner, as shown in the bottom panel of figure~\ref{fig5:timevar_specpar}. Taking into account the factors that 
 i) $r_{\rm in}$ is still kept constant, 
 ii) disk flux is proportional to $T_{\rm in}^{4}$, 
 iii) slightly harder spectrum $\Gamma\sim$1.9 than typical value of 2.0--2.2 in the HSS, 
 iv) high Compton fraction of $\sim$0.4, 
 and v) the flux corresponding to 0.4\% $L_{\rm E}$, 
 the BeppoSAX TOO6 data may have been taken during the state transition into the LHS.

 To investigate a possible effect by modeling of the hard Comptonized component on uncertainties for {\tt diskbb} parameters, we also tried to apply other thermal Comptonization model {\tt thcomp} (\cite{thcomp}) instead of {\tt simplcutx}, i.e. {\tt constant$*$TBabs$*$smedge$*$thcomp$\otimes$diskbb} (hereafter model B). The fits are slightly worse, but still reasonable giving the averaged reduced chi-square of 1.23 because electrons may have actually non-thermal distribution during the steep power-law phase. The time variation of $T_{\rm in}$ and $r_{\rm in}$ obtained from model B show a very similar trend to those from model A (figure \ref{fig5:timevar_specpar}), suggesting that modeling the hard  component does not have a significant effect on disk parameters.
 
\subsection{Correlation among X-ray spectral and timing properties}

\begin{figure}
\begin{center}
\includegraphics[width=8cm]{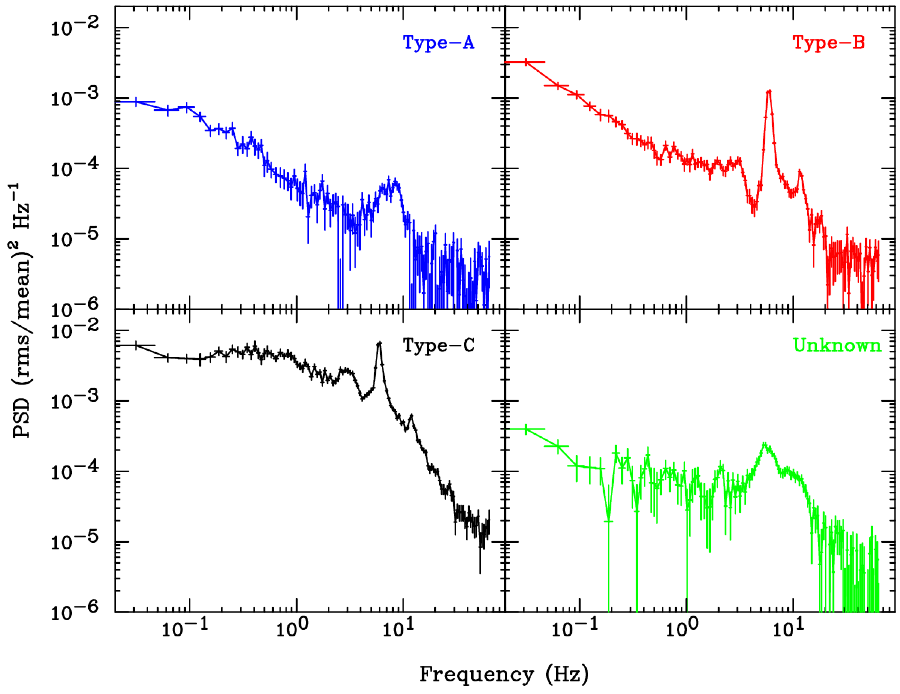} 
 \end{center}
\caption{Example of power spectral densities (PSDs) where Type-A (upper left), B (upper right), C (lower left), and unknown (lower right) QPOs appeared. The data are taken from RXTE/PCA observations with ObsIDs: 40124-01-12-00, 40122-01-01-03, 40124-01-11-00, and 40124-01-39-00, respectively.}\label{fig7:qpo_type}
\end{figure}

Here we study correlations among spectral parameters through our systematic analysis above, and compared them with timing properties.  We basically followed and used LF-QPO classification, such as Type-A, B and C, and QPO central frequency ($f_{\rm QPO}$) derived by \citet{qpodef2}, but we re-classified special case type C$^*$ into Type-C and B-Cathedral type into Type-B because they show similar characteristics in the rms-$f_{\rm QPO}$ plot [see figure 3 in \citet{qpodef2}]. For epochs where no QPOs are detected, there is no information about the fractional rms variability (hereafter we refer to rms) in \citet{qpodef2}, hence we calculated the rms in the range of 0.03--64 Hz using the PCA event, binned, or single-bit data with higher time resolution than the standard-2 data (16 sec). Figure~\ref{fig7:qpo_type} exhibits examples of PSDs. Four observations (obsID: 40124-01-13-00, 40124-01-14-00, 40122-01-01-00, and 40122-01-01-02) showed transitions from one Type to another during one pointing observation, so we performed the spectral analysis by separating each different QPO period. \citet{qpodef2} showed that Type-C QPOs appear in the rising phase before MJD~51479 and Type-A QPO is detected during very limited periods only on MJD~51467-51470, just around the first jet ejection (see section~3.4).

\begin{figure*}
 \begin{center}
\includegraphics[width=8cm]{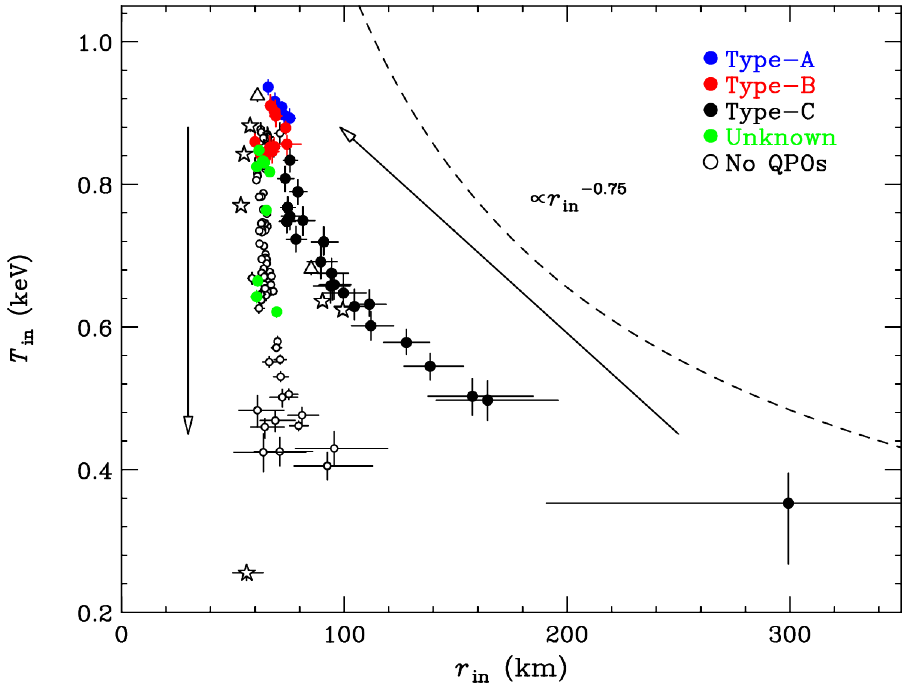} 
\includegraphics[width=8cm]{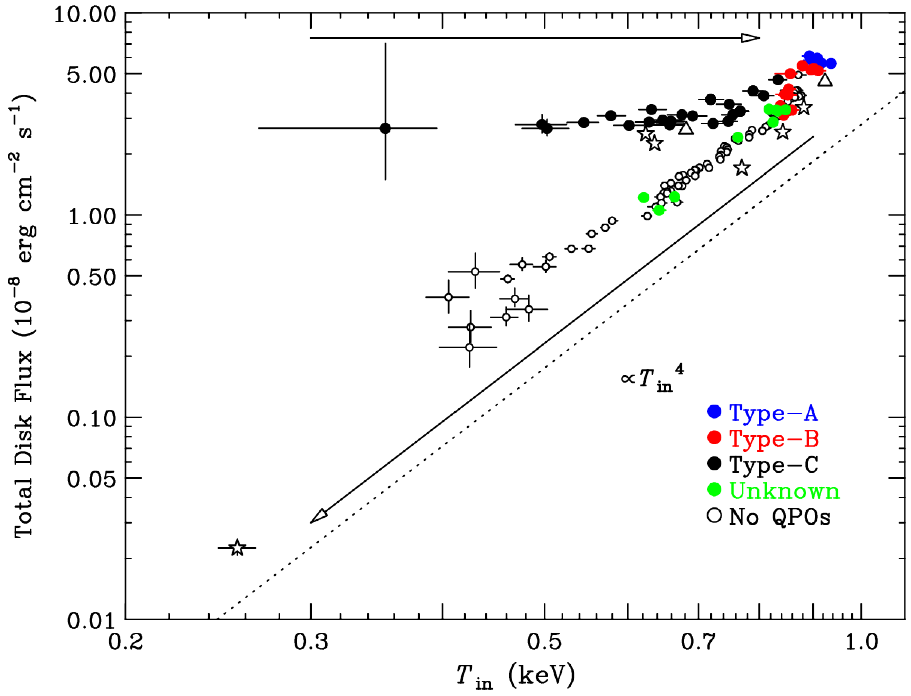} 
\includegraphics[width=8cm]{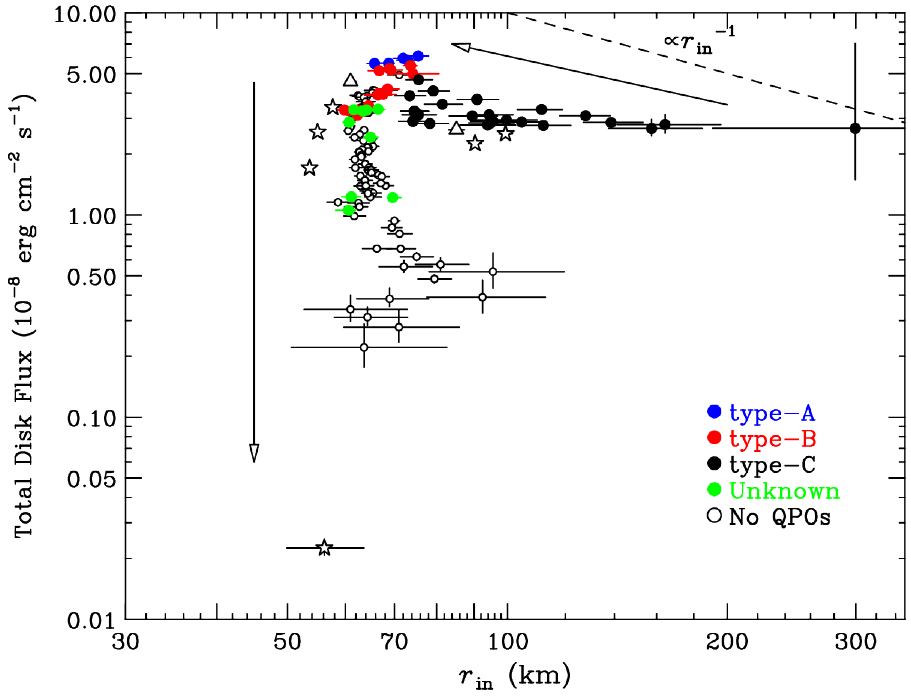} 
\includegraphics[width=8cm]{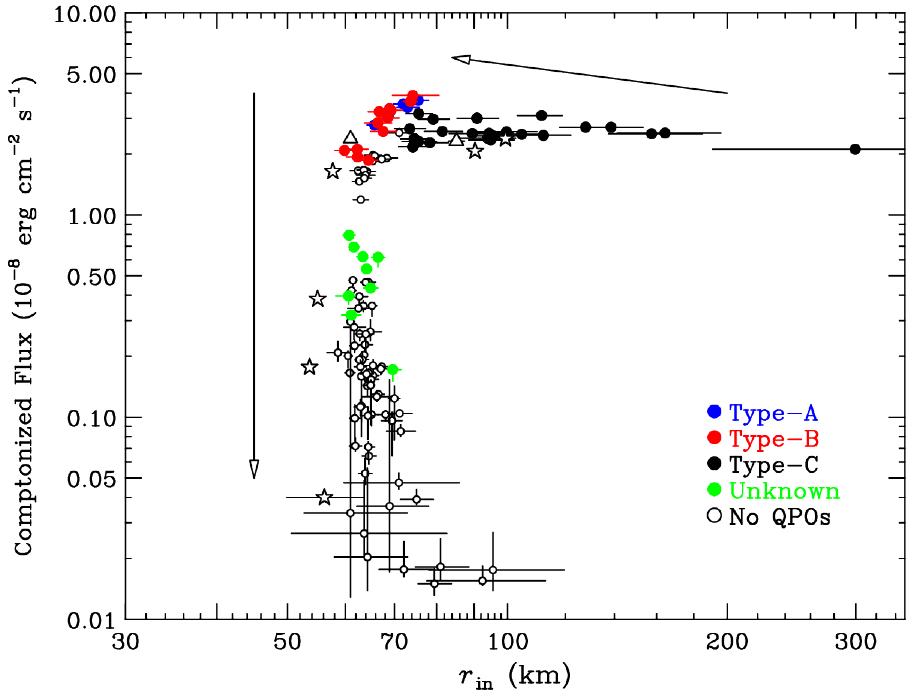}
 \end{center}
\caption{Upper left panel: Relation between innermost radius ($r_{\rm in}$) and temperature ($T_{\rm in}$). 
Two arrows show the time evolution; the right arrow for the rapid rising phase just before the peak flux, and the left arrow for the gradual decaying phase after the peak flux. Dashed curve indicates a relation of $\propto r^{-0.75}_{\rm in}$ as expected from the standard accretion disk for illustration purpose. 
 Total disk flux against $T_{\rm in}$ (upper right panel) and $r_{\rm in}$ (lower left panel). "Total" means that Compton up-scattered disk photons are included. Lower right panel: relation between $r_{\rm in}$ and Comptpnized flux. 
The points are classified into the five categories where Type-A (blue). B (red) and C QPOs (black), unknown (green), and no QPOs (open circles). The meanings of the stars and triangles are the same as those in figure~\ref{fig5:timevar_specpar}. Dotted line (upper-right panel) indicates the relation expected for a variable accretion rate with a fixed $r_{\rm in}$, whereas dashed line in lower-right panel again means the relation for a variable $r_{\rm in}$ with a constant accretion rate. 
}\label{fig8:rin_tin_diskflux}
\end{figure*}

Top panel of figure \ref{fig8:rin_tin_diskflux} shows relation between $r_{\rm in}$ and $T_{\rm in}$ in the MCD model. As can be seen from this figure, there are two separate branches: the constant $r_{\rm in}$ branch (vertically distributed in this plot) and the variable $r_{\rm in}$ branch. The variable $r_{\rm in}$ branch is composed of the epochs showing Type-C QPOs (including the rising phase), and conversely all the epochs with Type-C QPOs belong to the variable radius branch. In other words, Type-C QPOs appear only when $r_{\rm in}$ deviates from ISCO. The data in this branch are well fit by the relation of $T_{\rm in}({\rm keV})=(9.5\pm1.9)~[r_{\rm in}({\rm km})]^{-0.58\pm0.04}$, which resembles a relation expected from the assumption of the MCD model for $T_{\rm in}$ to be related to $r_{\rm in}^{-0.75}$ for a fixed accretion rate. The difference is likely due to a gradual reduction of the accretion rate during the $r_{\rm in}$ shrinking phase (lower-left panel of figure~\ref{fig8:rin_tin_diskflux} and corresponding text). 

It seems that the epochs with Type-C QPOs never reach $r_{\rm in} =$ ISCO, i.e. the occurrence of Type-C QPOs seems to cease just before $r_{\rm in}$ reaches the ISCO. This will be a key for figuring out the generation mechanism of Type-C QPOs.  For instance, this would  be naturally explained if the presence of a hot accretion flow (e.g., ADAF) inside an outer cool disk is essential for the generation of Type-C QPOs, such as trapped oscillations of a perturbation at the transition region (\cite{KatoManmoto2000}) or the LT 
precession (e.g., \cite{Ingram2009}). Type-A and B can not be clearly separated with each other in this plot, but both are seen just around the cross point between the two branches, during periods of almost maximum $T_{\rm in}$ at 0.8--0.9 keV and constant $r_{\rm in}$. 

The total disk bolometric flux against $r_{\rm in}$ and $T_{\rm in}$ also show the two distinct branches. We confirmed that the disk flux is proportional to $T^{4}_{\rm in}$, except for the epochs showing Type-C QPOs, as already suggested by observations (\cite{1655_spec,1550_spec,ldisk_tin}) and as expected from theoretically for the standard accretion disk with a fixed $r_{\rm in}$. There is no breaks at higher luminosity and temperature that are commonly seen in ultra-luminous state (ULS) of bright BHXBs; e.g., GRO J1655--40 (\cite{1655_spec}) and XTE J1550--564 (\cite{1550_spec}). The last data point, BeppoSAX TOO6 data on MJD 51629 when the disk component is not well constrained by RXTE data, is located at the extrapolation of the flux-$r_{\rm in}$ relation seen in RXTE data. Including the BeppoSAX TOO6 data, the flux- or luminosity-temperature relation ($\propto T^{4}_{\rm in}$) holds in the wide range over 0.26 keV to 0.94 keV. The accretion rate, in proportion to the total disk flux for a given $r_{\rm in}$, seems to decrease by $\sim 2.5$~dex during the decay phase over $\sim 150$~days.
   
However, during periods when Type-C QPOs are seen, this relation is violated, and the disk flux varies with the relation of $r^{-1}_{\rm in}$ and becomes almost flat at 3$\times$10$^{-8}$ erg cm$^{-2}$ s$^{-1}$, corresponding to 0.24 $L_{\rm E}$. It is reminiscent of the disk luminosity being inversely proportional to $r_{\rm in}$ for a fixed accretion rate $\dot{M}$ in the standard accretion disk model (disk flux $\propto \dot{M}/r_{\rm in}$). These results suggest that the mass accretion rate, which has been already increased before these observations start, is kept constant during the rising phase of the outburst (i.e., Type-C epochs). The fact that the HIDs start from the upper-right corner (figure~\ref{fig2:hid}), rather than the lower-right part, also supports that the accretion rate has already been increased before the observations begin. To be precise, the disk flux to $r_{\rm in}$ relation in Type-C phase is slightly flatter than $r^{-1}_{\rm in}$, indicating that the accretion rate decreases by a factor of $\sim 3$ (in $\sim 10$~days) as $r_{\rm in}$ approaches to the ISCO.

At the epochs with the maximum total luminosity 
($\approx 0.62 L_{\rm E}$; section 3.2), the direct disk luminosity is comparable to that of the Comptonized component (figures~\ref{fig2:hid} and
 \ref{fig5:timevar_specpar}), and $\approx 0.33 L_{\rm E}$. 
Considering the photons used for the Comptonized component, the disk 
luminosity at these epochs is about $0.48 L_{\rm E}$. 
Provided 
that $r_{\rm in}$ has reached to the ISCO and the disk accretion rate $\dot{M}$ has reduced by a factor of $\sim 3$ (i.e., down to $\sim 0.48$ times the Eddington accretion rate) by these epochs, the accretion rate at the beginning of these observations 
(MJD~51462) seems 
%
slightly ($\sim 1.5$ times) super-Eddington. 
If $r_{\rm in}$ after its shrinkage 
is $3 R_{\rm S}$ 
(with $R_{\rm S}$ being the Schwarzschild radius), then 
$r_{\rm in}$ at the beginning, which is $\sim \times 5$ larger value 
(figure~\ref{fig8:rin_tin_diskflux}), is about $15 R_{\rm S}$.
Another characteristic radius is 
the photon trapping radius $R_{\rm trap}$ [\cite{Begelman1978}; 
$R_{\rm trap} \approx 8 (\dot{M}/\dot{M}_{\rm E}) R_{\rm S}$,
where the Eddington accretion rate $\dot{M}_{\rm E}$ (shining 
at $L_{\rm E}$) is assumed to 
be 16~($L_{\rm E}/c^2$)]. 
Given that $R_{\rm trap} \approx 12 R_{\rm S}$ at the beginning 
(with $\dot{M}/\dot{M}_{\rm E} \approx 1.5$) and 
$\approx 4 R_{\rm S}$ when $r_{\rm in}$ reaches the ISCO 
(with $\dot{M}/\dot{M}_{\rm E} \approx 0.48$), 
$r_{\rm in}$ is decreasing roughly 
in keeping with $R_{\rm trap}$. 
Although various disk quantities (e.g., the temperature and the rotational 
velocity) deviate from the standard-disk formulae 
at $\ll R_{\rm trap}$ (\cite{Kawaguchi2003}), 
the disk quantities 
at $\gtrsim R_{\rm trap}$
can be described by the standard-disk formulae even for 
super-Eddington accretion rates (\cite{Kawaguchi2004}). 
Therefore, curves and lines in figure~\ref{fig8:rin_tin_diskflux}, 
which are expected for a standard disk, are applicable for the 
whole data in this work.

\begin{figure*}
 \begin{center}
\includegraphics[width=8cm]{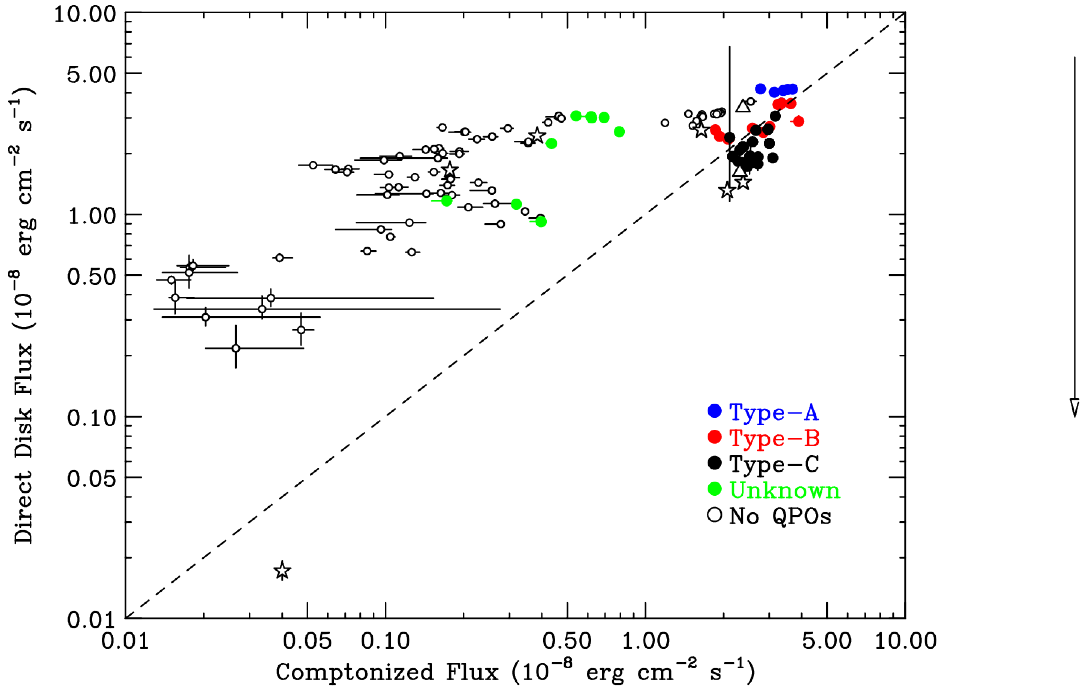}

\includegraphics[width=8cm]{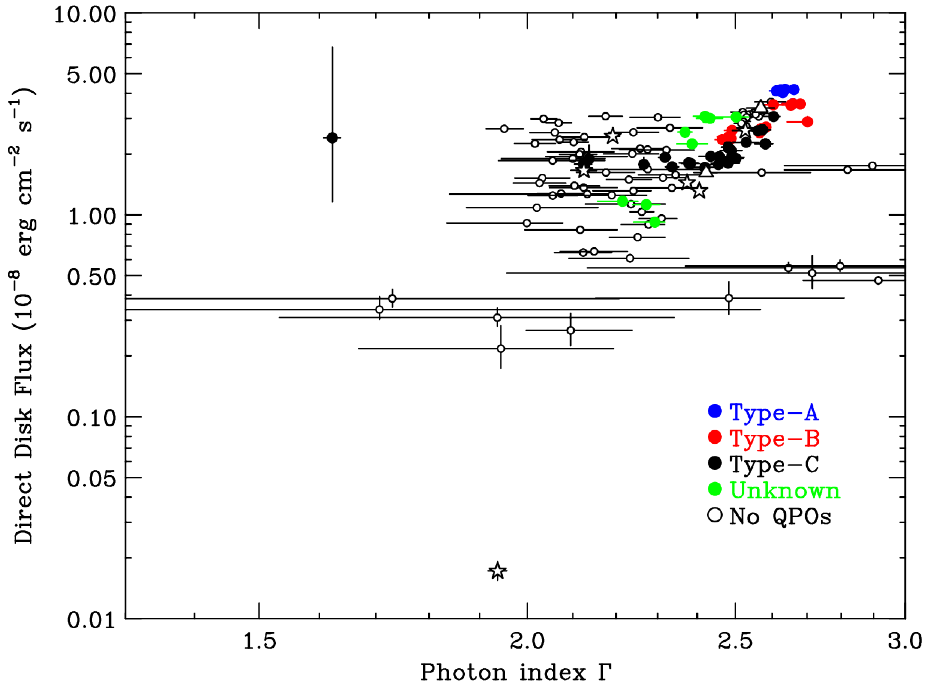} 
\includegraphics[width=8cm]{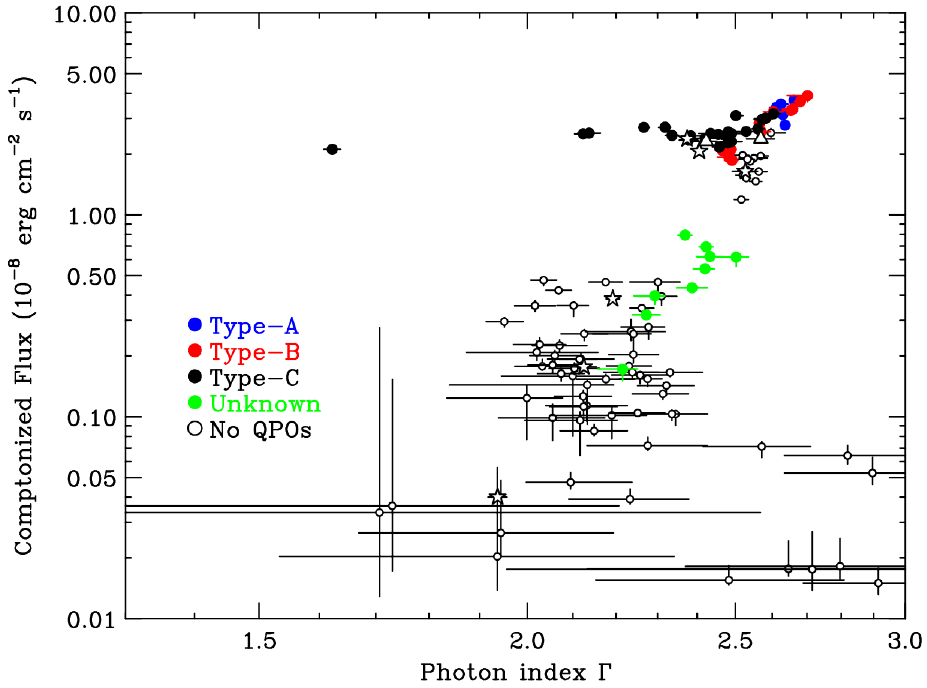} 
 \end{center}
\caption{Relation between Comptonized flux and direct disk flux, together with the dashed line where Comptonized flux equals to direct disk flux (upper panel). Relation of direct disk flux (lower left) and Comptonized flux (lower right) as a function of photon index ($\Gamma$). The meanings of the symbols are the same as those in figure~\ref{fig8:rin_tin_diskflux}.
}\label{fig9:phindex_diskhardflux}
\end{figure*}

In order to investigate the relations (i) between the properties of the hot medium ($\Gamma$, or Compton-$y$ parameter [=$4{\rm k}T_{\rm e}/(m_e c^2)\max(\tau,\tau^2)$ where k$T_{\rm e}$ is the electron temperature and $\tau$ is the optical depth] and QPOs, and (ii) between the presence/absence of QPOs and spectral states, we show the disk flux (direct disk component) and the Comptonized flux (scattered component) as a function of 
$\Gamma$ in figure \ref{fig9:phindex_diskhardflux}.  The QPO only appears when the Comptonized flux is higher than 2.0$\times$10$^{-8}$ erg cm$^{-2}$ s$^{-1}$ corresponding to 0.16 $L_{\rm E}$. Namely, the emergence (or the onset) of QPOs are closely related to the power of the hot Comptonizing medium. On the other hand, the disk flux does not show any clear cut-off for the presence/absence of QPOs. These results will provide useful constraints on plausible QPO models. During the Type-B QPO epochs, both the disk and the Comptonized fluxes correlate with $\Gamma$, as seen in other BHXBs (\cite{study_typeb}). As the system becomes brighter, the power-law component gets softer (with larger $\Gamma$).

\begin{figure}
 \begin{center}
\includegraphics[width=8cm]{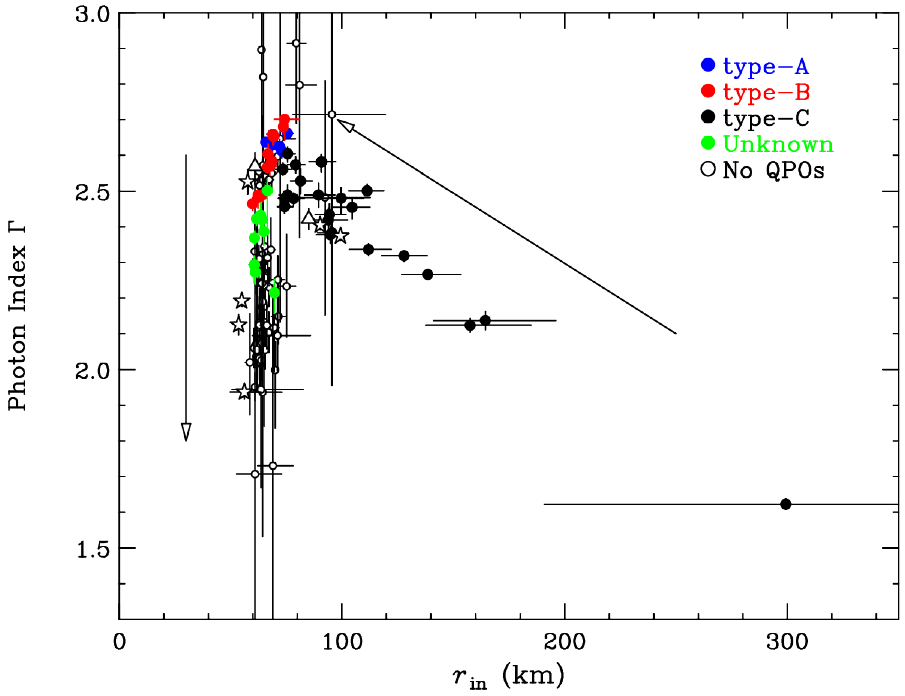} 
\includegraphics[width=8cm]{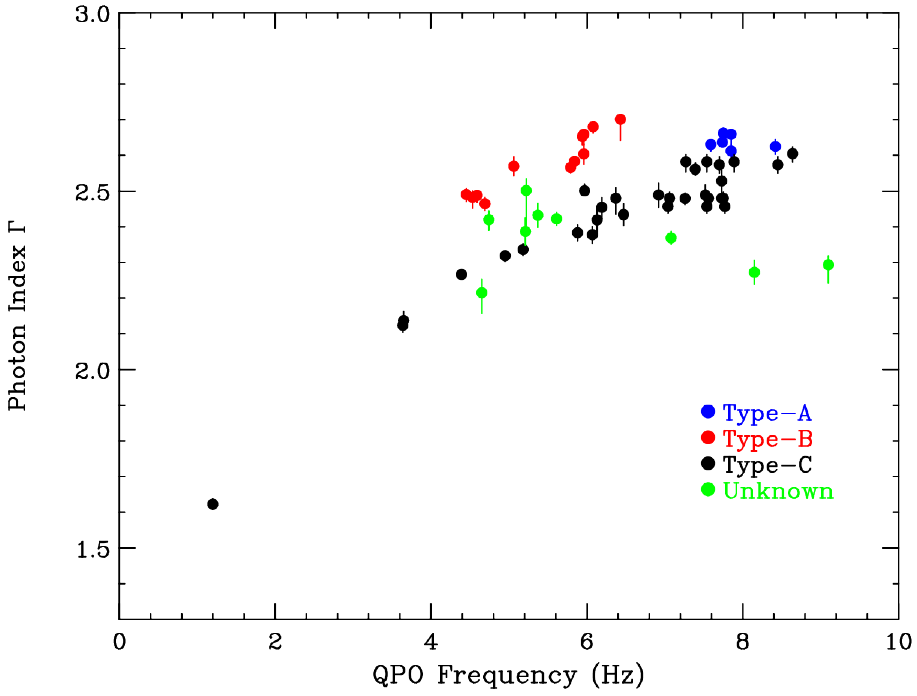} 
\includegraphics[width=8cm]{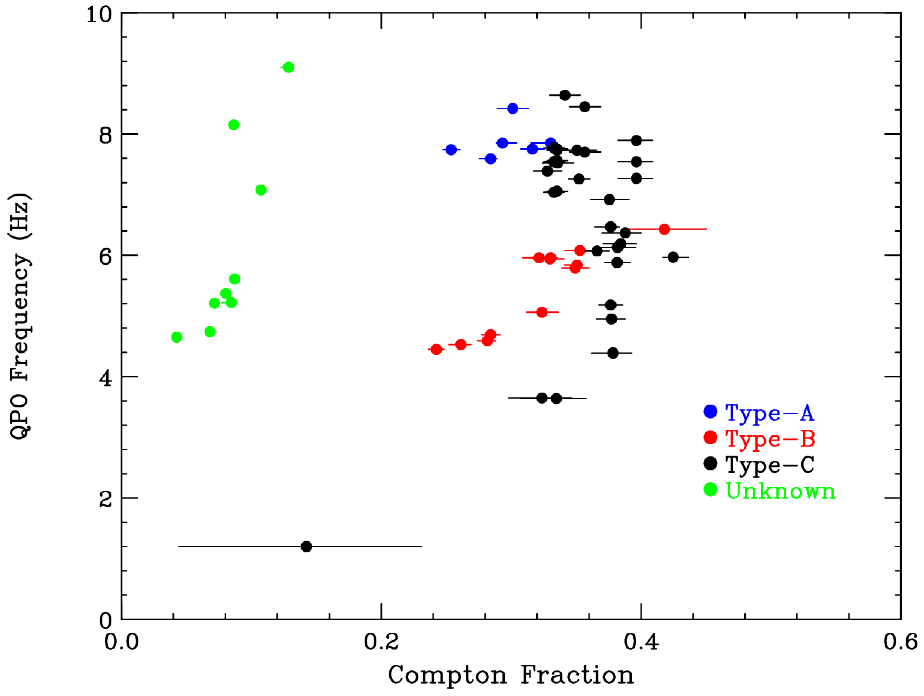}
 \end{center}
\caption{Top panel: Relation between innermost radius ($r_{\rm in}$) and photon index ($\Gamma$). Arrows indicate time evolution. Middle panel: Relation between QPO frequency ($f_{\rm QPO}$) and $\Gamma$. Bottom panel: Relation between Compton fraction $f_{\rm sc}$ in the {\tt simplcutx} model and $f_{\rm QPO}$. The same symbols as figure \ref{fig8:rin_tin_diskflux} are used in this figure. }\label{fig10:phindex_freq_rin}
\end{figure}

Moreover, $\Gamma$ v.s.\ $r_{\rm in}$ relation is plotted in figure \ref{fig10:phindex_freq_rin} to figure out how $\Gamma$ (related to the Compton-$y$ parameter and the degree of Compton cooling at the corona/ADAF) is affected by $r_{\rm in}$ and the types of QPOs. The $\Gamma$ increases from 1.6 to 2.6 in the Type-C QPO phase as $r_{\rm in}$ approaches to the ISCO and saturates at around 2.7, then gets slightly harder to $\Gamma \approx 2.0$ with $r_{\rm in}$ kept constant in the decaying phase. The correlation between $\Gamma$ and $f_{\rm QPO}$ in Type-C QPOs, shown in the middle panel, were also found for many 
BHXBs (\cite{index_freq, index_freq2,Zhang2024}).
Type-B epochs tend to show slightly smaller $\Gamma$ (slightly harder power-law) than Type-A epochs, which is visible in the middle panel of figure~\ref{fig10:phindex_freq_rin}. This may suggest a larger energy dissipation in the corona relative to that in the disk (such as stronger evaporation from the disk to the corona) in Type-B phases than Type-A. This will be mentioned in section~4.1, when a possible geometry is discussed. $\Gamma$ is correlated with $f_{\rm QPO}$ during Type-B QPO epochs. Type-A and -B phases show similar values for the Compton fraction ($f_{\rm sc}$; bottom panel), implying that the coronal optical thickness would be similar in the two QPO epochs. The four Types of QPOs including unknown type locate at distinct regions in this plot, indicating their different physical origins and/or different sites in the disk-corona-ADAF system. This could shed light on the geometry/configuration of the system and properties of the hot flow (e.g., optical thickness, etc.).

\begin{figure*}
\begin{center}
\includegraphics[width=8cm]{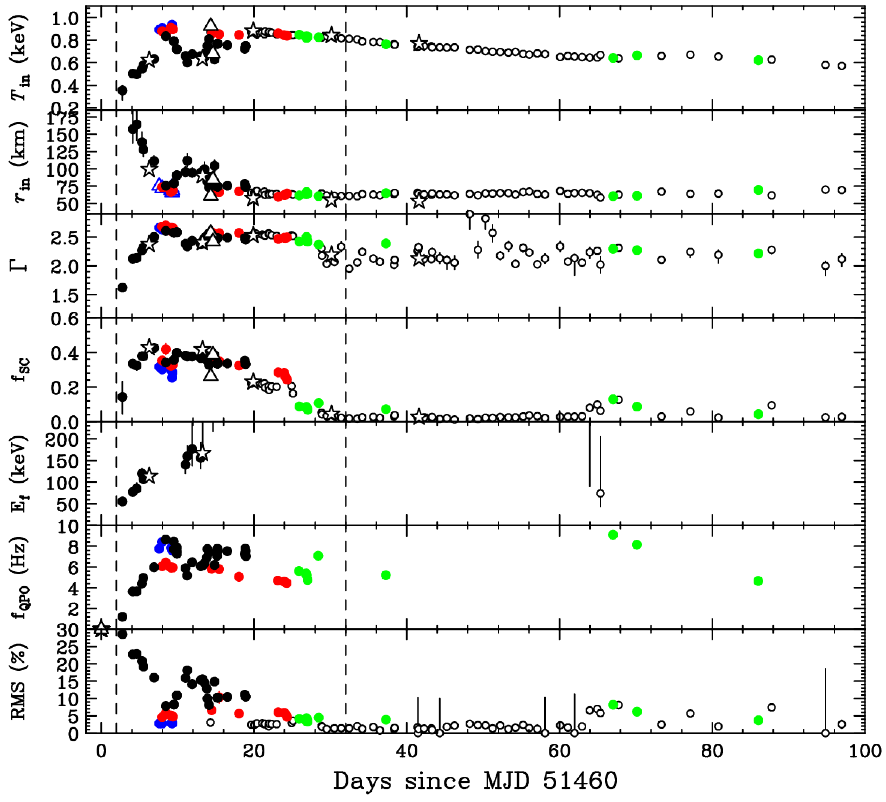} 
\includegraphics[width=8cm]{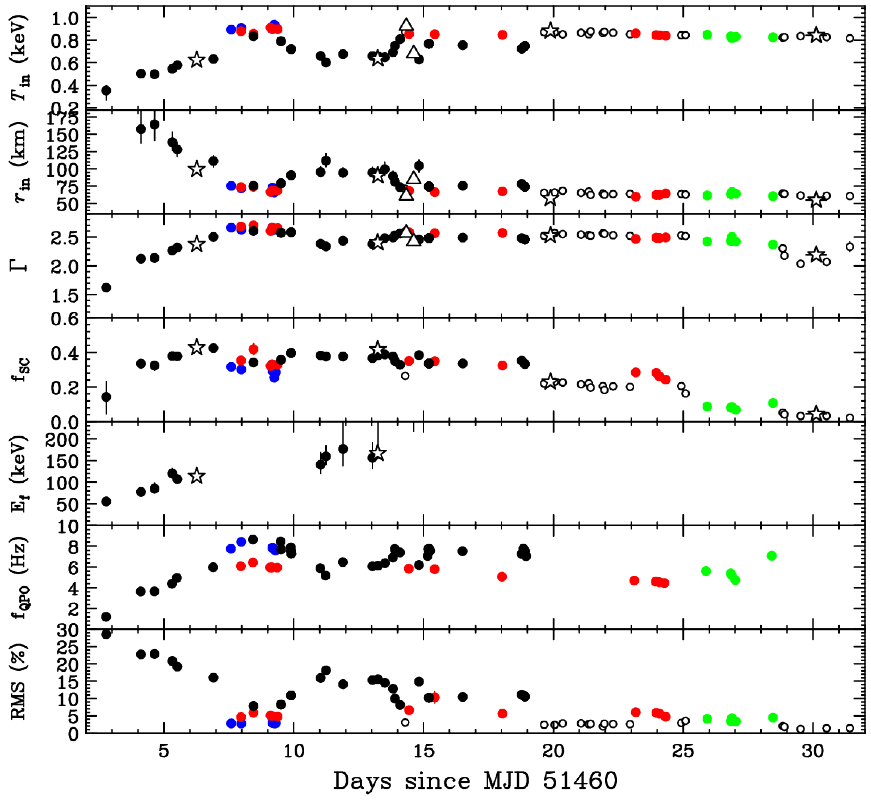} 
 \end{center}
\caption{Time evolution of spectral and timing properties. Innermost temperature ($T_{\rm in}$), innermost radius ($r_{\rm in}$), QPO frequency ($f_{\rm QPO}$), and fractional rms variability in 0.03--64 Hz are shown from top to bottom. Right panel: Zoom-up of Days 2--32 (indicated by dashed lines in left panel), where X-ray and radio data are investigated in figure \ref{fig13:relation_jet}. }\label{fig11:timevar_spectiming}
\end{figure*}

In figure \ref{fig11:timevar_spectiming}, we plot time variation of the spectral parameters together with timing parameters, $T_{\rm in}$, $r_{\rm in}$, $\Gamma$, $f_{\rm QPO}$, and rms.  Since rms used in this figure is calculated for the  0.03--64~Hz range, it is strongly correlated to the QPO rms (\cite{qpodef2}). These plots show how the QPO strength is affected by the accretion properties ($r_{\rm in}$, $\Gamma$, relative strength of Comptonizing medium, etc.).  The most remarkable result is that the time variation of $r_{\rm in}$ displays quite similar trend to that of the rms, despite independent analyses from spectral and timing. Especially during the rising phase, $r_{\rm in}$, the SHR, and the rms behaviour are well anti-correlated with $T_{\rm in}$, $\Gamma$, and $f_{\rm QPO}$ behaviour.
  
\begin{figure*}
 \begin{center}
\includegraphics[width=8cm]{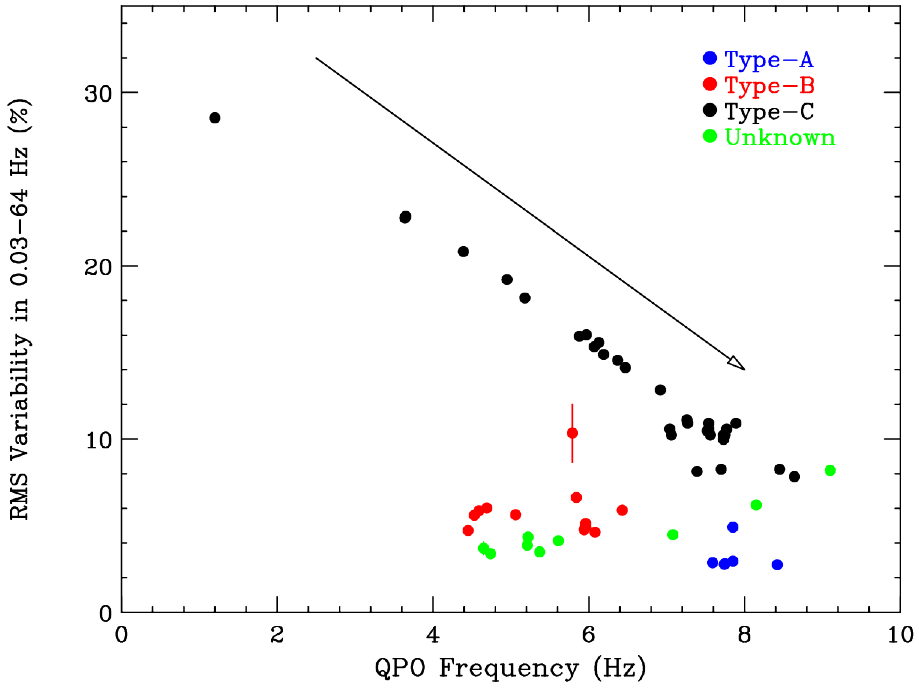}
\includegraphics[width=8cm]{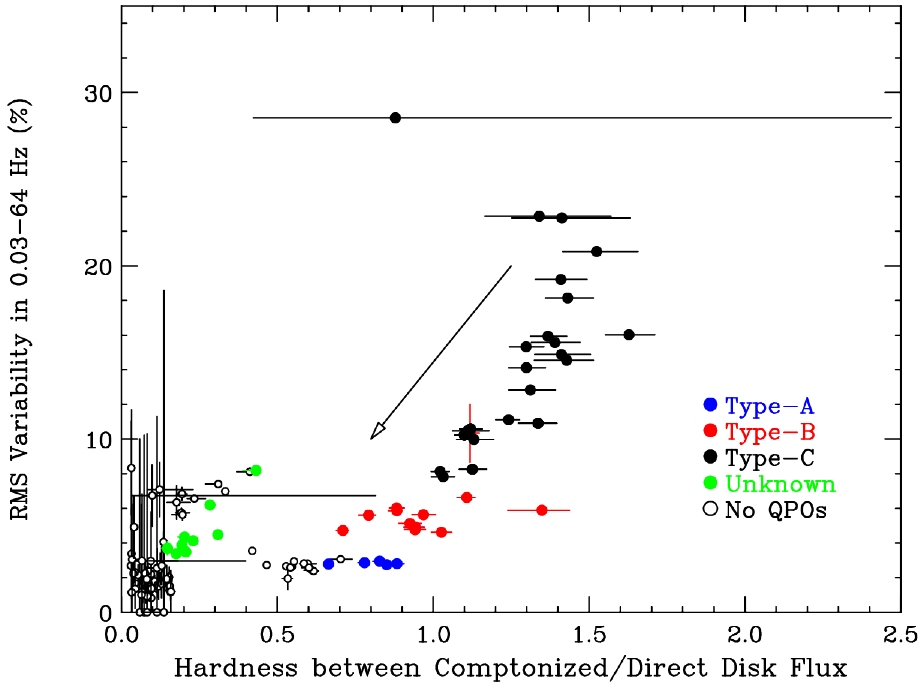}
\includegraphics[width=8cm]{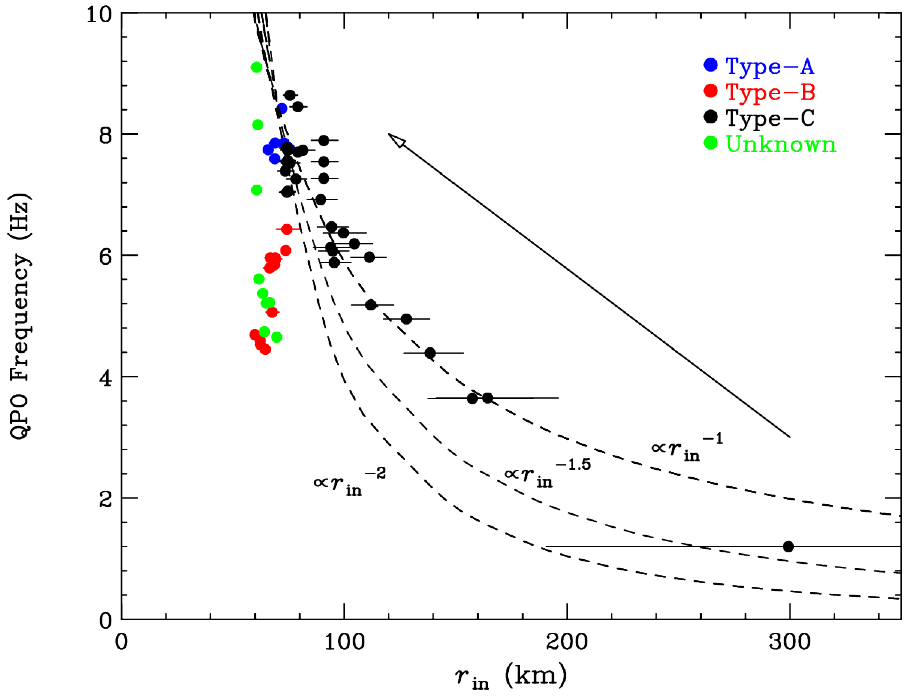} 
\includegraphics[width=8cm]{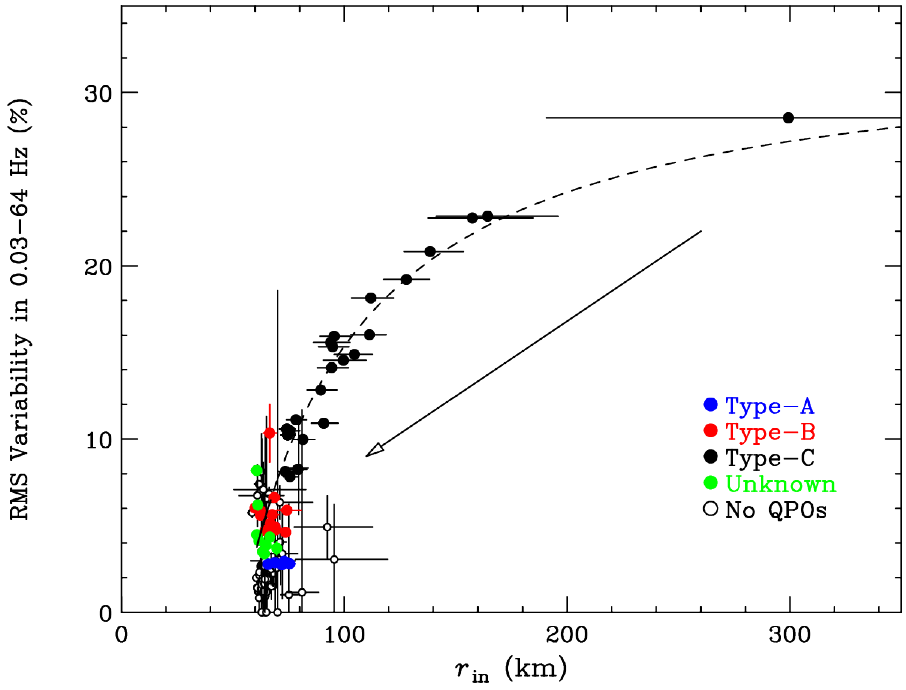} 
 \end{center}
\caption{Top left panel: Relation between rms and QPO frequency ($f_{\rm QPO}$), which is the same as figure 3 in \citet{qpodef2}. Top right panel: Relation between rms and spectral hardness ratio (SHR) of the Comptonized flux relative to the direct disk flux.
Lower-left panel: Relation between innermost radius ($r_{\rm in}$) and $f_{\rm QPO}$. Three dashed curves starting at $r_{\rm in}$ of 70~km and frequency of 8.5~Hz show $r^{-1.0}_{\rm in}$, $r^{-1.5}_{\rm in}$ and $r^{-2.0}_{\rm in}$ relation, for illustration purpose. A relation of frequency $\propto r^{-1.5}_{\rm in}$ is expected for dynamical/viscous timescales at $r_{\rm in}$.
Lower-right panel: Relation between innermost radius ($r_{\rm in}$) and fractional rms variability in the 0.03--64 Hz range. Dashed curve indicates the relation: rms[\%] = $33.0 - 25.3(r_{\rm in}/70[{\rm km}])^{-1.01}$,  combined from the best-fit parameters of rms-$f_{\rm QPO}$ (upper left) and $r_{\rm in}$-$f_{\rm QPO}$ (lower left) relation. 
In all the panels, arrows indicate the time evolution of the rapid rising phase.  
}
\label{fig12:rin_freq_rms}
\end{figure*}

To clarify the relations among these parameters, we show correlations in figure \ref{fig12:rin_freq_rms}.   
As also shown by \cite{qpodef2},
the three types of QPOs (A, B, and C) locate at distinct regions in the rms--frequency plot (with gaps in rms and/or in the frequency), and Type-A/B do not locate at the extrapolation of the Type-C QPO relations, suggesting that at least Type-A/B and Type-C arise from 
different physical origins. 
Indeed, \citet{qpo_geometry} argues a jet-based origin for Type-B QPOs from their dependence on viewing angles (stronger for more face-on views),
although Type-A QPO is not included in their results and discussions. 
 %
%

During the Type-C QPO periods, including the rising phase with decreasing $r_{\rm in}$, $f_{\rm QPO}$ is well anti-correlated with $r_{\rm in}$. This fact indicates that Type-C QPO is closely related to 
$r_{\rm in}$ (e.g., oscillations at the disk-ADAF transition radius; \cite{KatoManmoto2000}). The dynamical/viscous timescale in an accretion flow show a relation between characteristic frequencies and radial distance in the form  
$r^{-1.5}$ (\cite{textbook}). 
Negative correlation between $f_{\rm QPO}$ and $r_{\rm in}$ were also reported for other BHXBs (\cite{maxi1659,Misra2020,Zhang2024}), albeit larger scatter. For example, \citet{maxi1659} showed $r_{\rm in} \propto f_{\rm QPO}^{-0.70\pm0.12}$ for MAXI J1659--152, and \citet{Misra2020} showed a roughly linear relation in the $f_{\rm QPO}$--$r_{\rm in}$ diagram over a factor of $\sim$1.9 variation in $r_{\rm in}$ for GRS 1915+105, whereas we investigate 
over a wider $r_{\rm in}$ range (a factor of $\sim$5) in this work. \citet{Zhang2024} showed 
that $f_{\rm QPO}$ varies as roughly $\propto r_{\rm in}^{-1.6}$ for GX339--4, while they showed the scattered relation for EXO 1846--031.  
Although for LT models, 
relativistic precession of the inner accretion flow (ADAF), 
for Type-C QPOs also show anti-correlation between $f_{\rm QPO}$ and $r_{\rm in}$, 
\citet{Zhang2024} pointed out that the LT model does not explain the 
$f_{\rm QPO}$--$r_{\rm in}$ relation for the two BHXBs they analyzed. 
LT models 
would expect that $f_{\rm QPO}$ depends on $r_{\rm in}$ slightly more strongly (e.g., $f_{\rm QPO}\propto r^{-1.9}_{\rm in}$ from figure 5 in \cite{Ingram2009} and figure 1 of \cite{Veledina2013}). 
Thus, accurate determination of the $f_{\rm QPO}$--$r_{\rm in}$ relation for a number of objects turns out to be an efficient tool to discriminate plausible mechanisms of (Type-C) QPOs.

The larger rms for the larger $r_{\rm in}$ indicates that the flux time variation and QPOs are mainly caused by the hot accretion flow (e.g., ADAF or/and corona above the disk), rather than by the accretion disk. 
This can be confirmed by plotting rms as a function of SHR shown in right upper panel of figure~\ref{fig12:rin_freq_rms}. In the first half of the Type-C QPO phase (e.g., upper six 
data points), in which $r_{\rm in}$ is drastically decreasing, the covering factor of the disk as seen from the ADAF/corona increases, then leading to more seed photons for Comptonizing medium and to higher SHR. In the latter half of the Type-C phase, together with Type-A/B phases, data points in the rms--SHR plot are smoothly connected (towards smaller rms at smaller SHR) from Type-C, B, A, to no QPOs. Indeed, the spectral component responsible for QPOs is the Comptonized power-law component, rather than the disk component (e.g., \cite{Sobolewska2006}). Any plausible theory for Type-C QPOs also needs to reproduce a larger variability amplitude for more inclined (closer to an edge-on view) systems (\cite{qpo_geometry}). 

At an early stage of state transitions, Type-C QPOs usually appear, with $f_{\rm QPO}$s increasing and the variability amplitude decreasing as the X-ray spectrum softens (e.g., \cite{diskjet_maxij1820}). In this work, we found that the increase of the frequency is associated with the shrinkage of $r_{\rm in}$. It also turned out that the spectral softening (associated with the frequency increasing phase of Type-C QPO) is related to the increase of both $T_{\rm in}$ (figure~\ref{fig8:rin_tin_diskflux}) and $\Gamma$ (softening of the power-law component; figure~\ref{fig10:phindex_freq_rin}). 
In contrast, $f_{\rm QPO}$ in the Type-A/B QPO periods 
shows no clear relation with $r_{\rm in}$, which is kept nearly constant around the ISCO. Type-B QPO shows various frequencies at 4.5--6.5 Hz keeping similar $r_{\rm in}$, while Type-A QPO takes higher $f_{\rm QPO}$ at 7.5--8.5 Hz than Type-B. 

Since rms is anti-correlated with $f_{\rm QPO}$ in the Type-C phase, we can also see that there is a clear correlation between rms and $r_{\rm in}$. Plausible models for Type-C QPOs need to reproduce a larger amplitude rms for a larger $r_{\rm in}$, whereas no models so far seem to predict this behaviour. The linear relation between rms and $f_{\rm QPO}$ can be expressed as: rms[\%] = $33.0 - 3.0 f_{\rm QPO}[{\rm Hz}$]. Since  
$f_{\rm QPO}$[Hz] is fitted as $(8.45\pm0.13) (r_{\rm in} / 70 [{\rm km}])^{-1.01\pm0.07}$, the two relations reduce to rms[\%] = $33.0-25.3 (r_{\rm in}/70 [{\rm km}])^{-1.01}$. If this equation more or less represents the properties of the Type C QPO, it means that rms will saturate at a certain level ($\sim$33\,\%) as $r_{\rm in}$ increases. 
Type A and B QPOs, which occur with similar $r_{\rm in}$, show different rms with Type-A showing slightly lower rms than Type-B. 


\subsection{Correlation between X-ray and radio data}

Although it has been established that plasma are ejected as large-scale relativistic jets in association with the spectral state transition from 
 LHS to HSS in BHXBs, the problems upon when, where, and how they launch 
  still remain unresolved up to present. XTE J1859+226 during the 1999---2000 outburst displayed at least five jet ejection events identified by \citet{radio}, so 
   we may get some clues for this important questions from spectral and timing information. \citet{radio} estimated (and showed by arrows in their figure 4) rough peak times of radio flux density at Day 8, 14, 19, 23, and 27 since MJD 51460. The peak radio flux density was gradually decreasing with time, 110 mJy for the first flare, 20--40 mJy for the second and third flares, a few--10 mJy level for the fourth and fifth flares. 
   
\begin{figure*}
 \begin{center}
\includegraphics[width=14cm]{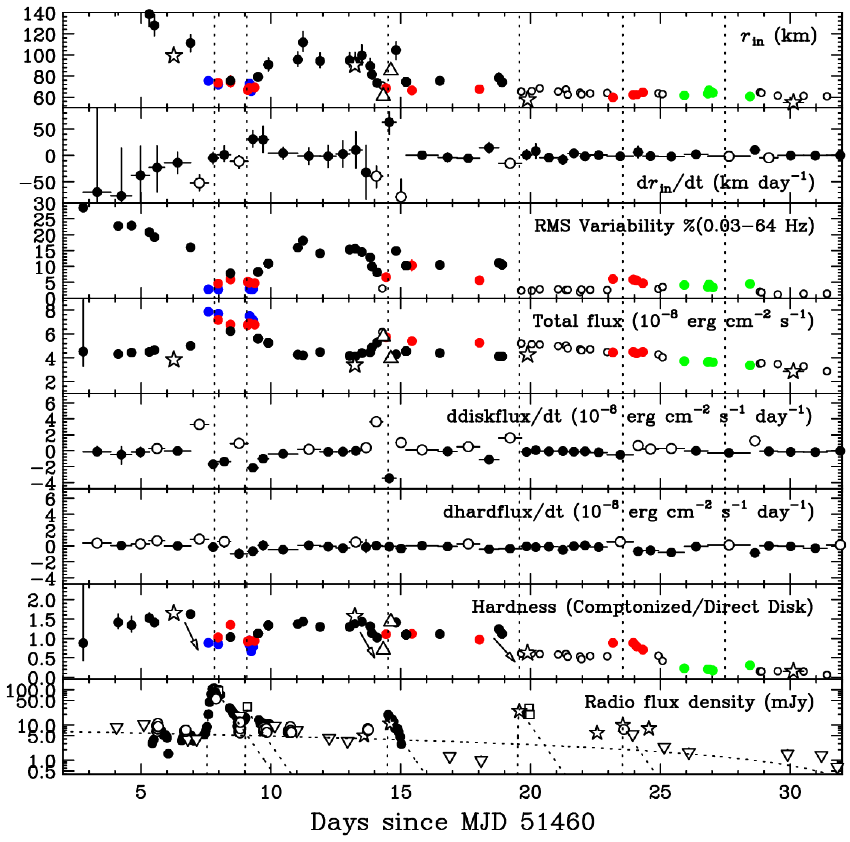} 
 \end{center}
\caption{Time variation of radio flux densities with spectral and timing properties. The innermost radius ($r_{\rm in}$), time derivative of $r_{\rm in}$, rms variability, total flux, time derivatives of direct disk flux and Comptonized flux, hardness ratio of Comptonized flux relative to direct disk flux, and radio flux density are shown from top to bottom. In the second panel, negative derivatives with their error-bars detached from zero are indicated by open circles, while in the fifth and sixth panels, positive values are by open circles.  
In the seventh panel, small three arrows indicate the 
steep drop of the hardness ratio ahead of the first three radio flares (dotted lines).
In the bottom panel, different symbols are used for different radio telescopes (MERLIN in filled circles, Ryle in open circles, GBI in open squares, RATAN-600 in open stars, and VLA in inverted triangles). Six dotted vertical lines indicate radio peak times estimated from the best-fit model shown in table \ref{tab3:fit_radioflux} and the timing of the fifth jet identified by \citet{radio}. Dotted lines indicate the components of fitting for the radio light curve.  In other plots, the meanings of the symbols are the same as those in figures ~\ref{fig8:rin_tin_diskflux}--\ref{fig12:rin_freq_rms}. }\label{fig13:relation_jet}
\end{figure*}

    To evaluate radio jet properties quantitatively, 
    we combine the radio data collected from VLA (1.43 GHz),
    MERLIN (1.66 GHz), GBI (2.25 GHz), and RATAN-600 (3.9 GHz), and fit the radio light curve with four brightening features (with their peak times at Day $\sim$7.8, $\sim$14.5, $\sim$19.6, and $\sim$23.6). 
They correspond to the four jet ejection events (except for the fifth jet event with much weaker fluxes than others) identified by \citet{radio}. We also found additional bump-like feature at Day $\sim$9.1 after the main peak of the first jet. Therefore, we consider this feature as another possible jet ejection and included this feature in the fitting of the light curve. For each feature, we apply the Fast Rise Exponential Decay (FRED) model with a linearly rising epoch till the peak time, 
followed by an exponential decay after the peak time (e.g., \cite{diskjet_maxij1820}).
The rising rate and the exponential decay timescale are assumed to be the same for all the five FRED components in the fitting, resulting in 393.2~[mJy day$^{-1}$] and 0.412~[day], respectively.
The light curve, till Day 35, is then fitted in a least-square manner 
by a combination of a linear and five FRED components, as shown in figure~\ref{fig13:relation_jet}.
The fitted peak times and peak fluxes for the five features 
are summarized in table~\ref{tab3:fit_radioflux}. 
For the second to the fourth jet ejections, radio measurements are 
sparse in time, and thus a wide range of fitting is possible.
Here, we chose the fitting results giving the smallest peak fluxes for them.
For instance, the rising timescale (the peak flux divided by the rising rate) is $\sim 0.28$~day and $\sim 0.10$~day for the first and the third jet ejections.

Figure \ref{fig13:relation_jet} shows time variation of radio flux density together with spectral 
    and timing properties with vertical dotted lines at the five peak times plus one possible bump shown above. The radio flares including the possible bump of the first jet coincide with the timing that the $r_{\rm in}$ decreases toward the ISCO or that it has already reached the ISCO. 
    We also calculated time derivatives of $r_{\rm in}$, the direct disk flux and the Comptonized flux, by connecting two data points separating more than 0.3 days with a straight line to grasp trends of various quantities in $\gtrsim$~sub-day timescales.
    Open circles indicate the epochs when the time derivatives 
    ($\pm$~error-bars) of $r_{\rm in}$ (or the fluxes) are definitely negative (or positive). There is a hint that the negative $dr_{\rm in} / dt$ (i.e., a rapid shrinkage of $r_{\rm in}$) seems to be associated with the jet ejections. Drops of rms associating with radio activities within a few days, reported in the literature (e.g., \cite{jet2}), are seen also in this figure.
   In addition, the radio flux density during each flare tends to be larger as the periods (duration), during which $r_{\rm in}$ has been deviated from the ISCO till each radio flare, are longer. 
We will discuss these points in section~4.3 in more detail. 

We summarize the relations between the jet production and various types of QPOs as follows. As described in section~3.3,  
Type-A QPOs appear only around the first jet ejection (figure~\ref{fig13:relation_jet}). Therefore, Type-A QPO may not be a necessary condition for launching a jet. Since Type-B QPOs appear around the peaks of the first, the second, and the fourth jets (and before the third jet), a connection 
between Type-B and jet launching is inferred (\cite{Soleri2008, jet,jet2,diskjet_h1743,diskjet_maxi1535}). \citet{diskjet_maxij1820} indeed showed, with dense cadence for another BHXB MAXI J1820+070, that Type-B QPO emerges just before the jet ejection. Type-C appears through the third jet (Day 19.6). After Type-C QPO disappears or after the deviation of $r_{\rm in}$ from the ISCO becomes small, the power of a newly launched jet (an enhancement in radio emission) may be small.

\begin{table}
{\footnotesize
\caption{Fitting results of radio light curve with a linear plus 5 FRED components. } \label{tab3:fit_radioflux}
\begin{tabular}{ccccc}\\\hline
Jet      &  Peak time\footnotemark[$*$] & Peak flux & Rising rate\footnotemark[$\dagger$] & Decay time\footnotemark[$\dagger$] \\        
         &   (Day)      & (mJy)     & (mJy/day)    &  (day)     \\\hline
1st      & 7.82$\pm$0.01 &  111.1$\pm$3.2   & 393.2$\pm$24.5 & 0.412$\pm$0.010 \\     
1st bump & 9.08$\pm$0.01  & 26.21$\pm$2.40   &   --- & --- \\
2nd      & 14.52(fixed)  & 13.09$\pm$1.09   &   --- & --- \\	
3rd      & 19.57(fixed)  &  28.28$\pm$9.35   &   --- & --- \\	
4th      & 23.56(fixed)  &    8.87$\pm$4.19   &   --- & --- \\	
\hline
\end{tabular}
\begin{tabnote}
\footnotemark[$*$] Days since MJD 51460.\\
\footnotemark[$\dagger$]Both parameters are set in common for the five transient jets. \\
Errors are quoted at statistical 68\%. In the fitting, the 5th flare peak at Day 27.5 suggested by \citet{radio} is not included due to its much weaker flux 
 density than other flares.\\
\end{tabnote}
}
\end{table}

\section{Discussions}
%

\subsection{Accretion Disk Geometry}
We applied spectral modeling with the MCD model from a geometrically-thick, optically-thin accretion disk, assuming that a part of disk photons are up-scattered due to Compton scattering in the hot medium. 
The observational results, especially the time evolution of the outburst, can be naturally and quantitatively explained in terms of the following sandwich geometry between disk and corona/ADAF geometry.
\begin{enumerate}
\item At very beginning of the outburst around the discovery, the source 
 hard X-ray flux above 20 keV increases very rapidly within one day and reaches almost the maximum accretion rate when RXTE observations starts.
 The slightly lower energy cutoff at $\sim$50 keV corresponding to electron temperature of $\sim$20 keV (in contrast to the nominal value 50--100 keV in electron temperature; e.g., \cite{cygx1_broadband}) is a result of effective Compton cooling in the corona/ADAF from the disk photons. 
\item During the initial rising phase, the mass accretion rate is kept almost constant. Later, $r_{\rm in}$ shrinking leads to a disk flux increases, while conversely the rms variability amplitude, i.e. a ratio of variable hard component relative to stable disk component, decrease. The increase of the folding energy $E_{\rm f}$ which reflects the electron temperature (k$T_{\rm e}$) during the rising phase (figure~\ref{fig5:timevar_specpar}) implies that the Compton cooling due to soft photons from the cold disk is ineffective at that phase. For instance, the optical depth $\tau$ (or the density) of the corona/ADAF may decrease as $r_{\rm in}$ approaches the ISCO and compensate the increase of the seed photons from the disk due to the shrinkage of $r_{\rm in}$. Then, the reduction of the optical depth $\tau$ would lead to the softening of the power-law component (i.e., smaller Compton-$y$ parameter and thus larger $\Gamma$). Another scenario is that the Comptonizing plasma consists of mixed and hybrid non-thermal and thermal electrons, and non-thermal electrons are becoming more dominant in the hybrid plasma. In this case, $\Gamma$ is directly related to an index of the original power-law energy distribution of the electrons. Type-C QPO originates from the disk-oscillation around the truncation radius $r_{\rm in}$, hence $f_{\rm QPO}$ increases as the $r_{\rm in}$ decreases.
 \item During the decay phase, $r_{\rm in}$ stays at the ISCO. In this phase, the mass accretion rate decreases with  $r_{\rm in}$ kept constant. The Comptonized component changes independently of the disk component, and Type-B QPOs sometimes appear in this phase when the hard component gets brighter than a certain luminosity.  
\end{enumerate}

Our disk/corona geometry is different from one assumed in the Lamppost model (e.g., \cite{lamppost}). While $r_{\rm in}$ reaches the ISCO in the SIMS evolved from large truncated $r_{\rm in}$ in the LHS in the disk-ADAF geometry (\cite{ADAF2}), $r_{\rm in}$ already reaches the ISCO in the bright LHS in the Lamppost model. During the state transition from the LHS to the HSS, the size of the jet-like corona above the disk changes and gets smaller in the Lamppost model (\cite{lamppost_nicer,lamppost_nicer2}). Although the direction of the corona is different from each other, i.e. inside 
in the truncated disk model and vertical 
(along 
the rotation axis) in the Lamppost model, shrinkage of the corona/ADAF is similar to each other. 

\begin{figure}
 \begin{center}
\includegraphics[width=6.5cm]{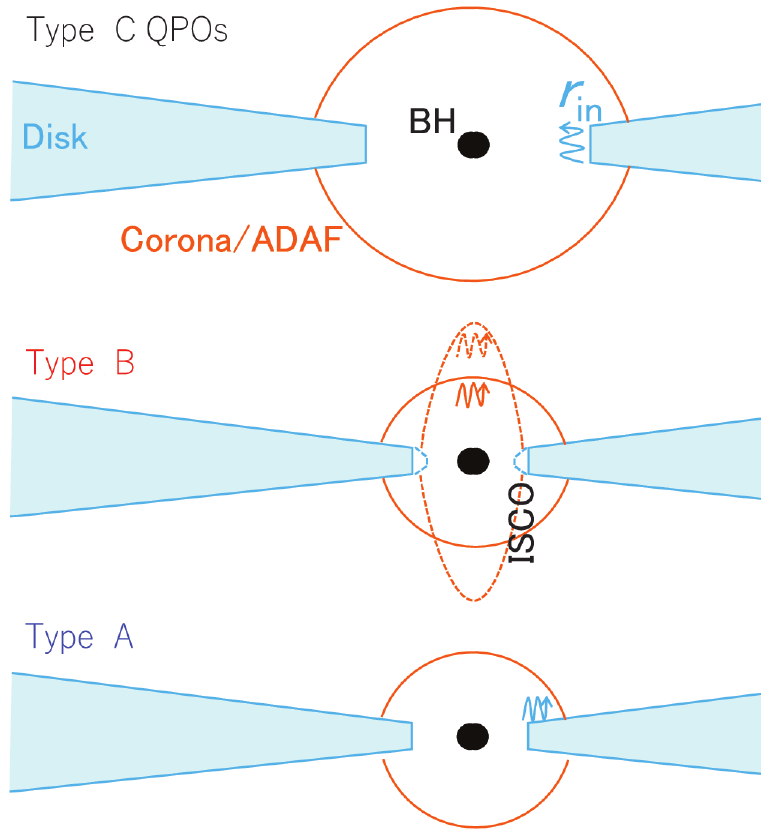} 
 \end{center}
\caption{Illustration for a possible configuration (see text in section~4.1 for details). A geometrically-thin (with a small aspect ratio) disk is surrounding the inner hot accretion flow (ADAF) in the Type-C QPO phase. In the Type-B phase, the corona above the disk is supposed to be unstable, generating a larger rms than the Type A phase. 
The height of the ADAF/corona is assumed to vary, resulting in a relation between $\Gamma$ and $f_{\rm QPO}$. 
When the energy ejection at the corona or the evaporation to the corona becomes small, a calm situation is realized with the smallest rms (Type A), fluctuating with the largest (among other Types) frequency on average.}\label{fig14:geometry}
\end{figure}

Based on the spectral and timing properties presented in section~3, a possible configuration/scenario, as a function of the QPO Types, is drawn in figure~\ref{fig14:geometry}. During epochs with Type-C QPOs, $r_{\rm in}$ is well detached from the ISCO (figure~\ref{fig8:rin_tin_diskflux}). In other words, Type-C QPOs stop before $r_{\rm in}$ reaches the ISCO. Perturbations at larger $r_{\rm in}$ will oscillate with lower frequencies (figure~\ref{fig12:rin_freq_rms}). 
When $r_{\rm in}$ is approaching to the ISCO rapid enough, in the process from Type-C to Type-B, a jet ejection will happen (section~4.3). For GRO J1655--40, Type-B and Type-C QPOs 
 were detected at the same time (\cite{1655_typebc}). This result is a direct evidence for different physical origins of Type-B and C QPOs. The co-existing situation may be archived at a marginal case of $r_{\rm in}\simeq$ISCO in our model. 

Type-A and B QPOs are generated only when $r_{\rm in}$ is very close to the ISCO (figure~\ref{fig8:rin_tin_diskflux}). Epochs with Type-B QPOs show slightly harder power-law emission than Type-A epochs (figures~\ref{fig9:phindex_diskhardflux} and \ref{fig10:phindex_freq_rin}). Since both QPOs do not coexist at the same time, a certain control parameter is required for switching between them. A relative energy dissipation rate in the corona compared to that in the disk, which is often referred as {\it f} in radiative transfer calculations of the disk-corona system around BHs (e.g., \cite{Haardt1991,Kawaguchi2001}), could be a candidate parameter. Similarly, a strength of the evaporation from the disk (\cite{Meyer1994,Liu2002}) can be an alternative. When the relative strength of the corona or the evaporation is intense enough, Type-B QPOs appear, and contrary, Type-A QPOs will emerge when they are small.

A positive correlation between the photon index $\Gamma$ and 
$f_{\rm QPO}$ (figure~\ref{fig10:phindex_freq_rin}) 
may mean that 
the height of the corna/ADAF could vary 
within the Type-B phase,  
as follows. 
A taller (more vertically inflated) corona, as a result of more energy release in the corona and/or 
a more intense evaporation from the disk, will suffer from a lesser 
Compton cooling (i.e., harder power-law with smaller $\Gamma$) by the disk seed photons, since the covering factor of the disk as seen from the corona is smaller than a less inflated corona.  
Oscillations taking place in a tall corona may result in a smaller frequency of QPOs, given that various timescales would have positive correlations with the distance from a BH. 
Incidentally, a time variation of the height of the hot Comptonizing medium within a spectral state is inferred for another BHXB MAXI~J1820+070 based on X-ray reverberation (\cite{lamppost_nicer}).

A relatively weaker corona or a less intense evaporation in the Type-A phase (compared to the Type-B phase) may be supported by the larger disk fluxes of the Type-A phase (figures~\ref{fig8:rin_tin_diskflux} and \ref{fig9:phindex_diskhardflux}). Although mechanisms of the Type-A QPOs, as well as those of Type-B, are still unclear, the disk-corona interface (as a remained location other than the disk at $r_{\rm in}$ and the corona) may be relevant to Type-A oscillations. For instance, the Kelvin-Helmholtz instability at the disk-corona interface (\cite{Shadmehri2010}), due to a difference of the radial velocity between them, could produce some fluctuations. When a vertical motion there (e.g., evaporation from the disk) becomes fast enough (entering to the Type-B phase), a difference of the radial velocity becomes a secondary factor, halting the generation of Type-A QPOs. The smaller rms of Type-A QPOs compared to that of Type-B (figure~\ref{fig12:rin_freq_rms}) will constrain plausible models of both oscillations.

\subsection{Black Hole Mass and Spin Estimation from the Continuum Fitting}

There are several methods for the BH spin estimation using X-ray data; Continuum Fitting Method (CFM; \cite{CFM1,simpl2}), X-ray Reflection Spectroscopy (XRS; \cite{xrs1,xrs2}), and Relativistic Precession Model (RPM; \cite{LT2,1655_RPM}). However, they are still in progress, because all these methods are not consistent with each other, e.g., see table 1 in \citet{1859_xrs}. Hence, we need to perform careful study of these methods. 

In this paper, we apply the CFM taking into account uncertainties of distance, inclination angle, and BH mass. We found that the $r_{\rm in}$s were kept constant after the X-ray peak flux in spite of the large flux variation.
The constant $r_{\rm in}$ is probably due to the presence of the ISCO, which corresponds to 3$R_{\rm S}$ for a Schwartzchild BH with the spin parameter $a_{*}=0$ and 0.5$R_{\rm S}$ for a maximum spinning Kerr BH ($a_*\simeq$1), see e.g., \citet{CFM1}. 
Hence, we can constrain the BH mass and spin using measured $r_{\rm in}$. 

By fitting the data whose Compton fraction $f_{\rm SC}$ is less than 0.20 (\cite{simpl2}),  the constant radius was derived to be 64.2$\pm$0.2 km.  However, this estimation for $r_{\rm in}$ is not taken into account inner boundary conditions close to the BH and spectral hardening due to electron scattering, so we need to correct for such effects following the method described in \citet{bhmass_est} for a non-rotating BH. 
\begin{equation}
M_{\rm BH}=\frac{c^2f^2\eta r_{\rm in}}{6G}
\end{equation}
where $c$ is the speed of light, $G$ is the gravity constant, $f$ is the spectral hardening factor which is typically 1.7 (\cite{hardening}), $\eta$ is the boundary condition factor (=0.412). We can derive the BH mass of 8.6$\pm$0.1 $M_{\odot}$, which is quite consistent with BH mass estimation of 7.8$\pm$1.9 $M_{\odot}$ measured by optical observations (\cite{opt_mass}). 
 Even if we use the BeppoSAX-measured $r_{\rm in}$=56.6$\pm$0.3~km, which shows smaller value by about 10\% than RXTE, we also get the consistent value of  7.6$\pm$0.1 $M_{\odot}$ with one measured in optical. However, we note that this estimation strongly depends on assumption of $D$=8 kpc and $i$=66$^{\circ}$.6. 

To confirm these results further, we also fitted spectra with another continuum model from accretion disk: {\tt slimbh} instead of {\tt diskbb}. This model takes into account the fully relativistic effect around a Kerr BH and is applicable to  wider range of luminosity up to the Eddignton limit than other model {\tt kerrbb2} (\cite{kerrbb2}). Actually \citet{appli_slimbh} applied {\tt slimbh} and {\tt kerrbb2} to the RXTE spectra of LMC X--3 over 0.1--1 $L_{\rm E}$, and derived fully consistent values of the spin parameter between both models. This model can implement effects by self-consistent vertical structure with non-LTE radiation transfer, calculated by the TLUSTY code\footnote{http://tlusty.oca.eu/index.html} if we do not use a fixed spectral hardening factor.  We fixed $M$=7.8 $M_{\odot}$, $D$=8 kpc, and $i$=66$^{\circ}$.6, and applied the following six (2$\times$3) cases: for viscous parameter $\alpha$=0.01 and 0.1, and for spectral hardening factors the TLUSTY calculation, and two fixed values of 1.6 and 1.7. We applied to only BeppoSAX TOO4 and TOO5 data because the disk component is dominant over the wide-band spectra in these two data (see figure \ref{fig4:ascaxtesax_fit}). The best-fit parameters are shown in table \ref{tab4:fit_slimbh}. The spin parameter $a_*$ (limited to be positive in this model) is constrained to $<$0.02 and 0.17$\pm$0.02 for $f$=1.7 and TLUSTY, and $f$=1.6, respectively. These results support a relatively low spin (similar to a Schwartzchild BH)
for this object in comparison with other BHXBs. 

However, taking into account uncertainties on  the distance and the BH mass, we found that $a_*$ is quite uncertain. $a_*$ can increase up to 0.7--0.8 in case of ($D$, $M$, $i$)=(5, 9.7, 62.3).  On the contrary,  assuming the BH parameters $M$=7.85 $M_{\odot}$ and $a_*$=0.149 derived from the RPM (\cite{LT_appli}), we can constrain $D$ to 7.2--8.3 kpc which is well within $D$=8$\pm$3 kpc assumed in this paper. 
If we change $a_*$=0.986 derived from the XRS method (\cite{1859_xrs}), the model prefers lower distance less than 5 kpc and lower luminosity than 0.05$L_{\rm E}$, but we could not get acceptable fits due to the limited luminosity range (0.05--1) $L_{\rm E}$.  
\begin{table}
{\footnotesize
\caption{Constraints on spin parameter and distance in {\tt slimbh} model fitting.}\label{tab4:fit_slimbh}
\begin{tabular}{ccccc}\\\hline
\multicolumn{5}{l}{Case-I:  $M$=7.8 $M_{\odot}$, $D$=8 kpc, $i$=66.6 deg. }\\\hline
Obs. & $\alpha$ & $f$\footnotemark[$*$] & $a_*$ & $\chi^2$/d.o.f \\\hline
TOO4 & 0.01 & 1.6 & 0.165$_{-0.007}^{+0.007}$ & 150.4/151\\
 &  & 1.7 & $<$0.001 & 187.1/151\\
 &  & --1 & $<$0.002 & 207.9/151\\
 & 0.1 & 1.6 & 0.166$_{-0.008}^{+0.007}$ & 150.0/151\\
 &  & 1.7 & $<$0.001 & 189.2/151\\
 &  & --1 & $<$0.001 & 284.2/151\\\hline
TOO5 & 0.01 & 1.6 & 0.173$_{-0.010}^{+0.010}$ & 141.6/146\\
 &  & 1.7 & $<$0.003 & 153.6/146\\
 &  & --1 & $<$0.016 & 163.1/146\\
 & 0.1 & 1.6 & 0.175$_{-0.010}^{+0.010}$ & 141.3/146\\
 &  & 1.7 & $<$0.004 & 152.6/146\\
 &  & --1 & $<$0.005 & 161.6/146\\
\hline
\multicolumn{5}{l}{Case-II:  $M$=7.85$M_{\odot}$, $i$=66.6 deg., $a_*$=0.149(fixed)} \\\hline
Obs. & $\alpha$ & $f$\footnotemark[$*$] & $D$ (kpc) & $\chi^2$/d.o.f \\\hline
TOO4 & 0.01 & 1.6 & 8.18$_{-0.06}^{+0.06}$ & 149.9/151 \\
 &  & 1.7 & 7.27$_{-0.05}^{+0.05}$ & 149.4/151 \\
 &  & --1 & 7.34$_{-0.04}^{+0.05}$ & 169.0/151 \\
 & 0.1 & 1.6 & 8.18$_{-0.06}^{+0.06}$ & 149.7/151 \\
 &  & 1.7 & 7.28$_{-0.05}^{+0.05}$ & 149.1/151 \\
 &  & --1 & 7.19$_{-0.04}^{+0.05}$ & 162.6/151 \\\hline
TOO5 & 0.01 & 1.6 & 8.25$_{-0.08}^{+0.08}$ & 141.2/146 \\
 &  & 1.7 & 7.32$_{-0.07}^{+0.07}$ & 141.0/146 \\
 &  & --1 & 7.59$_{-0.07}^{+0.06}$ & 162.7/146 \\
 & 0.1 & 1.6 & 8.26$_{-0.08}^{+0.08}$ & 141.0/146 \\
 &  & 1.7 & 7.34$_{-0.07}^{+0.07}$ & 140.6/146 \\
 &  & --1 & 7.48$_{-0.03}^{+0.03}$ & 155.8/146 \\
\hline
\end{tabular}
\begin{tabnote}
\footnotemark[$*$] --1 indicates model calculation based on TLUSTY code.\\
Errors are quoted at statistical 68\%.\\
\end{tabnote}
}
\end{table}

\begin{figure}
\begin{center}
\includegraphics[width=6.5cm]{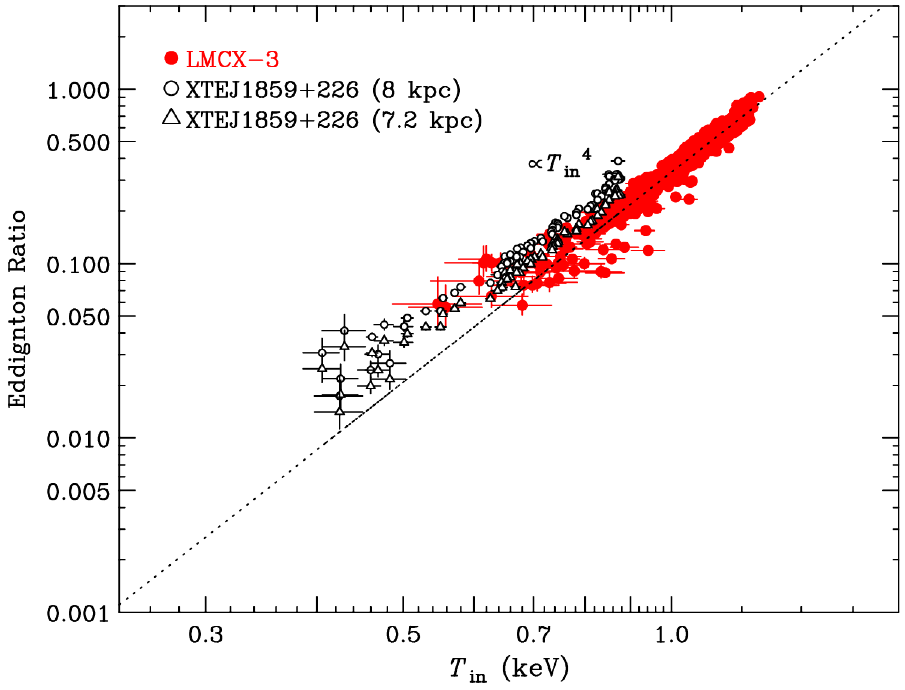} 
 \end{center}
\caption{Comparison of $T_{\rm in}$ and the total disk luminosity relative to the Eddignton luminosity between XTE J1859+226 [for two assumed distances: 8 kpc (shown by open circles) and 7.2 kpc (by open triangles)] and LMC X-3 (by red filled circles). For the latter, $T_{\rm in}$ is corrected for the small difference of the BH masses (see text). Dotted line shows the best-fit model of $T_{\rm in}^4$ for LMC X-3.}\label{fig15:comp_lmcx-3}
\end{figure}

In order to confirm validity of the source distance, we compared the disk-luminosity 
 properties with the BHXB LMC X-3 with well-known BH parameters 
  ($M$=6.98$\pm$0.56$M_{\odot}$, $i$=69.24$\pm$0.72 degrees; \cite{lmcx-3_mass}) 
  and distance $D$=49.59$\pm$0.09(stat.)$\pm$0.54(sys.)~kpc (\cite{lmcx-3_distance}). We analyzed 
 archival RXTE data of LMC X-3 in the same way as XTE J1859+226, and 
fitted RXTE/PCA spectra in the 3--20 keV range with the {\tt TBabs*simpl$\otimes$diskbb} model.
  The hydrogen column density $N_{\rm H}$ and photon index $\Gamma$ were fixed at 5.18$\times$10$^{20}$ cm$^{-2}$ (the Galactic value; \cite{nh}) and 2.2 (the nominal value in the HSS), respectively in the fitting. Taking into account the BH mass dependence on 
   $T_{\rm in}$, i.e., $T_{\rm in} \propto M^{-1/4}$ for a fixed Eddington ratio of the standard accretion disk model, $T_{\rm in}$ was corrected by a factor of $(7.8/6.98)^{-1/4}$ to adjust to the value of XTE J1859+226.  
  Figure~\ref{fig15:comp_lmcx-3} shows comparison of $T_{\rm in}$-$L_{\rm disk}/L_{\rm E}$ (i.e. Eddignton ratio of the disk luminosity) diagram between XTE J1859+226 and LMC X-3. 
  If the two sources have similar spin parameters (i.e., a relatively low spin parameter of LMC X-3; \cite{lmcx-3_spin}), they are expected to exhibit similar relations in this diagram. 
  For XTE J1859+226, only the data which show no QPOs are plotted and the source distance are assumed to be two cases: 7.2 and 8 kpc. We confirmed $L_{\rm disk}/L_{\rm E} \propto T_{\rm in}^4$ for both sources (\cite{1655_spec,ldisk_tin}) and found that they follow almost the same $T_{\rm in}-L_{\rm disk}/L_{\rm E}$ relation, and the $D$=7.2 kpc case was closer to that of LMC X-3. Although there are still uncertainties in the inclination angle $i$ and the BH mass $M$ for XTE J1859+226, $D$=8 kpc and slightly closer distance $D$<8 kpc would be a reasonable estimation.
  
To summarize this subsection, we found that both simple calculation based 
 on the $r_{\rm in}$ measurement and detailed spectral modeling with {\tt slimbh} support a relatively low spin for the assumed distance of $D$=8 kpc and inclination angle of $i$=66.6 degrees. The {\tt slimbh} model can also explain the observed X-ray continuum with the spin parameter $a_*$=0.149, which is obtained from the RPM, and distance of 7.2--8.3~kpc.  
 
\subsection{Jet Production}

Based on $r_{\rm in}$ and its time derivative, the jet production (the radio emission enhancement) seems to happen when fulfilling 
the following two conditions that $r_{\rm in}$ reaches the ISCO, and that $d r_{\rm in} / dt$ is negative. Albeit small number statistics, the negative $d r_{\rm in} / dt$ may precede the peaks of the radio flux density, although uncertainties on time-lags between a radio flare 
and a jet launching (\cite{Corbel2003,Corbel2005}) make further detailed discussion difficult.

The 
radio 
flares
seem to correlate with the increases 
of the disk flux (seen in the first-to-third jets), whereas they are not related to the Comptonized flux. 
It is more easily recognized in the time derivatives of the fluxes of the two components. 
Therefore, materials of the jet unlikely come from the hot gas flow (corona above the disk or ADAF) in this source. Probably, the infalling disk gas at $r_{\rm in}$, which is approaching to the ISCO, is the origin of the jet material. Detailed studies of jet and corona relation were done by \citet{jet_corona1} and \citet{jet_corona2} in several sources such as GRS 1915+105 and GX 339--4. Decreases in hard X-ray flux and in optical depth $\tau$ of the corona were identified prior to jet ejections [see figure 1 in \citet{jet_corona1} and figure 2 in \citet{jet_corona2}]. They suggest that the corona shrinks and is converted into jet materials.  Further investigations are needed for this issue. 

In order to confirm considerations above and to make clear 
what drives 
the jet ejections, 
we investigate the radio light curve along with the X-ray 
spectral fitting results. 
To investigate by which the five brightening features in section~3.4 were driven, 
we calculate a radio fluence (an integration of the flux) for each feature using the best-fit results in table \ref{tab3:fit_radioflux}. 
At an epoch between 1.28~day and 0.1~day before the peak time, 
the radio flux density of the FRED component is integrated over the coming 1-day interval from that epoch.
If there is more than one day ($\sim$timescale for rising plus decaying) gap 
in the radio light curve, a presence/absence of a radio activity is not sure. 
Then, the fluence is not assigned for such epochs.

\begin{figure*}
 \begin{center}
\includegraphics[width=14cm]{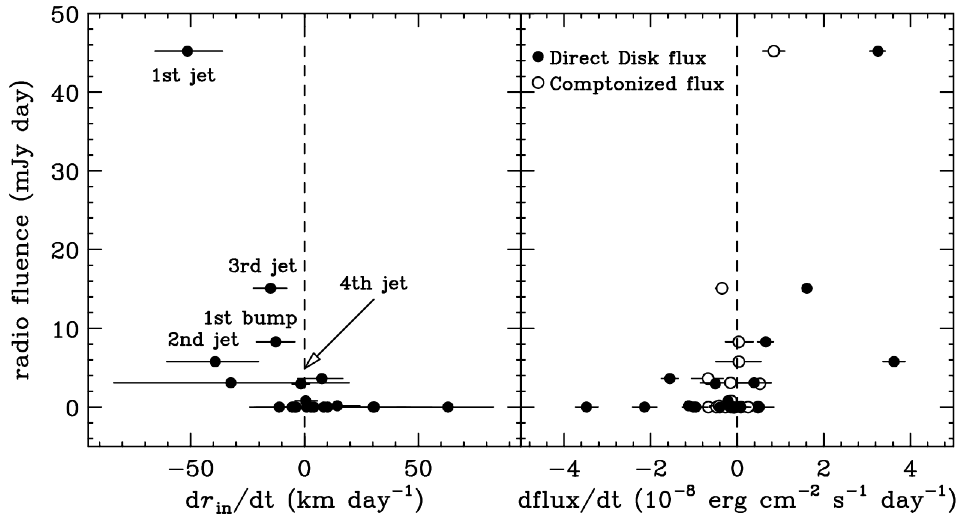} 
 \end{center}
 \caption{Relation between time derivatives of $r_{\rm in}$(left) and fluxes (right) with radio fluence estimated from the fitting. Direct disk flux and Comptonized flux are indicated by filled and open circles, respectively. 
 }
\label{fig15:radiofluence_relation}
 \end{figure*}

Figure~\ref{fig15:radiofluence_relation} shows the 
radio fluence, 
as a function of the time derivatives of $r_{\rm in}$, the direct 
disk flux and the Comptonized flux. 
As we discuss with figure~\ref{fig13:relation_jet}, the large radio 
flares
seem 
to be associated with a large negative $d r_{\rm in} / dt$ and a large 
increasing rate of the disk flux. 
Whereas the time variation 
of the Comptonized flux has lesser or no 
impact on the radio flares. 
Since jet ejections likely take place in much shorter timescales, 
radio measurements with denser cadence are certainly necessary. When $r_{\rm in}$ is decreasing rapidly  to the ISCO, a radio activity is expected to happen in the coming one day. Once a rapid decline of $r_{\rm in}$ is confirmed to be a precursor of jet ejections in a number of objects, it will provide various observational and theoretical perspectives. For instance, (i) it can be, as well as a drop of rms, a useful criteria  
to trigger ToO monitoring to catch the moment of the jet production 
in X-ray, radio and other facilities. 
It is more effective for sources with known ISCO radii via previous X-ray observations in their HSS.
Given the correlation between rms and $r_{\rm in}$ (figure~\ref{fig12:rin_freq_rms}), our discussions involving $d r_{\rm in} / dt$ provide a physical basis for the phenomenological fact that drops of rms are associated with jet ejections (e.g., \cite{jet2}). 
(ii) When $r_{\rm in}$ is already at the ISCO, jet ejections are not expected for a while until it leaves from the ISCO and gets ready for a next shrinkage of $r_{\rm in}$. (iii) It will also present conditions for jet-launching mechanisms in theoretical works. A model, which launches jets all the time when $r_{\rm in}$ is at ISCO, would have difficulties.

Since the "jet line" is likely related to the necessary condition for launching jets, radio flares 
are supposed to associate with downward passage across a certain hardness ratio. The time evolution of the hardness ratio in figure~\ref{fig13:relation_jet} (same as figure~\ref{fig5:timevar_specpar}) shows that the rapid declines of the hardness ratio indeed happen ahead of the first three radio flares. Although the downward passage of the hardness ratio [corresponding to the leftward jumps in the 
HIDs (figure~\ref{fig2:hid})] may suggest that the coronal gas forms the jet ejection, discussion above on the light curves for each of the disk and the Comptonized components indicates that it would not be the case.

In order to figure out where each jet event locates in the 
HIDs, we emphasized the pairs of two epochs sandwiching the 
timing of each jet event in figure~\ref{fig2:hid}. Here, we 
employ the Day $t_{\rm peak}-$0.28 to $t_{\rm peak}$ (with $t_{\rm peak}$ 
being 
the  peak time of the radio flare shown in 
table \ref{tab3:fit_radioflux}) as the timing of jet ejections. 
There is no single vertical line, over which 
all the pairs cross (e.g., \cite{jet2}). 
At somewhere within a "band" or a "belt" with non-zero width, 
rather than a line, jet ejections seem to happen accompanying 
leftward jumps in the HIDs. 
Then, it is difficult 
to predict the timing of jet-launching based solely on 
the position of a source in the HID. 
Since discussions in terms of $r_{\rm in}$ contrarily have a 
reference 
point (i.e., the ISCO radius), predictions of the timing 
(for, e.g., successful ToO observations) based on a 
physical basis can be made. 

\citet{jet} and \citet{jet2} 
suggested that the "jet line" originates from inward-moving of $r_{\rm in}$ to the ISCO, which is naturally expected to occur during the LHS-to-HSS transition, but direct observational evidences remained unclear for a long time. In this paper, we actually showed that radio flares occur when $r_{\rm in}$ rapidly moved to the ISCO quantitatively. Many authors showed a strong connection between jet ejections and a rapid drop in X-ray time variability and/or an appearance of Type-B QPO, but a unified physical picture about the jet ejections in BHXBs has not been established yet (e.g., \cite{jet2,diskjet_h1743,diskjet_maxi1535,diskjet_maxij1820}). Both phenomena 
are surely related to the ISCO, as we showed a clear correction between rms and $r_{\rm in}$ (figure~\ref{fig12:rin_freq_rms}), and an appearance of Type-B QPO 
when $r_{\rm in} \simeq$~ISCO (figure~\ref{fig8:rin_tin_diskflux}).  
If the inward-moving of $r_{\rm in}$ from a truncated radius to the ISCO is  the essence of the jet production, 
1) no ejections would happen among Type-A/B/Non-QPO transitions, and 2) not only Type-C to Type-B transitions but also Type-C to Type-A/Non-QPO transitions would be equally necessary conditions for jet-launching.

\section{Summary}

In this paper, we compiled available all the data covering soft X-rays to soft gamma-rays of the black hole binary XTE J1859+226 in the 1999--2000 outburst from RXTE, ASCA, BeppoSAX and CGRO, and performed systematic spectral analysis using the multi-color disk model convolved with the SIMPL Comptonization with an exponential high-energy cutoff. 
Based on this model, our findings are summarized as follows.
\begin{enumerate}
\item In the initial rising (of the flux at $\geqq 20$~keV) phase of the outburst, both innermost radius ($r_{\rm in}$) 
   and temperature ($T_{\rm in}$) drastically change with time. A high energy cutoff is surely present and evolves with time from 55~keV to 200~keV or more {in association with the shrinkage of $r_{\rm in}$}.
\item In the decaying phase after MJD 51480, $T_{\rm in}$ smoothly changes from 0.87~keV to 
  0.26~keV, keeping constant $r_{\rm in}$ at around 60 km. This is an evidence for 
    a presence of the innermost stable circular orbit (ISCO). 
\item Based on the disk flux v.s.\ $r_{\rm in}$ plot and the hardness-intensity diagrams, the gas accretion rate seems 
to have already been increased ($\sim$ 
the Eddington rate) when the observations start. 
Then, the accretion rate remains almost constant, or it 
gradually decreases slightly by a factor of $\sim$3, as $r_{\rm in}$ is shrinking towards the ISCO in the initial rising phase (giving the maximum total luminosity of $\sim$0.62 Eddington luminosity at the end of this phase).
\item In the rising phase, $r_{\rm in}$ repeated moving closer and farther away from the ISCO several times. Jet production might be related to this process. 
\item The photon index $\Gamma$ is correlated with the disk flux, the Comptonized flux, and $r_{\rm in}$.  
\item Type-C QPOs appear when $r_{\rm in}$ is located far away from the ISCO 
    in the rising phase, and its appearance ends just before $r_{\rm in}$ reaches the ISCO.  It implies that the presence of a hot flow inside an outer disk is essential for Type-C QPOs. We found a negative correlation between $f_{\rm QPO}$ and $r_{\rm in}$ ($f_{\rm QPO} \propto r^{-1.01}_{\rm in}$) from XTE J1859+226, as reported in earlier works on other BHXBs. All of these results seem to be inconsistent with the relation $r^{-1.9}$ expected from the Lense-Thirring precession. Thus, an accurate determination of the $f_{\rm QPO}-r_{\rm in}$ relation for a number of objects turns out to be potentially a powerful tool to discriminate plausible Type-C QPO mechanisms.
      In contrast, for the Type-A/B QPOs, $r_{\rm in}$ is very close to the ISCO and $f_{\rm QPO}$ changes independently keeping a constant $r_{\rm in}$. QPOs, whatever Types, appear only when the Comptonized 
flux exceeds a certain level, whereas their presence/onset 
do not show clear relation to the disk flux.
\item We found that the time evolution of $r_{\rm in}$ and the fractional rms variability in the 0.03--64 Hz show remarkably similar trend, exhibiting a tight correlation between the two parameters. The variability is determined by the balance of a stable cold disk and a highly variable ADAF/corona.
\item We applied relativistic accretion disk model {\tt slimbh} to the BeppoSAX data, and 
      found that the spin parameter is relatively low, adopting the distance of 8 kpc and the BH mass of 7.8$M_{\odot}$.
      If we assume the BH system parameters estimated from QPO Lense-Thirring effect, the distance is constrained to 7.2--8.3 kpc. 
\item We suggest that timing of radio flares coincides with the rapidly shrinking $r_{\rm in}$ from a large truncated radius to the ISCO, and at this timing the disk flux rapidly increases and the hardness ratio gets softer with little variation of the hard Comptonized component. This fact suggests that the jet materials originate from accreting matter from the innermost part of the disk rather than the ADAF/corona in this source.  
Furthermore, a rapid decline of $r_{\rm in}$ down to the ISCO (as well as a drop of rms)  seems to be a precursor of radio brightening events, and is likely an useful index for triggering ToO monitoring observations to catch the moment of the jet production. 
Compared to the association of rms drops with radio activities mentioned in earlier works, the time derivative of $r_{\rm in}$ likely describes the processes of jets more physically. 
It will also provide useful constraints on theories for launching jets.
\end{enumerate}

Both X-ray spectral and timing observations are very important to understand the 
 accretion disk states which can change very rapidly. 
 Dense, simultaneous coverage with 
  radio observations are also crucial to understand disk-jet interactions.
  Future Neutron star Interior Composition Explorer (NICER) and enhanced X-ray Timing and Polarimetry Mission (eXTP) observations together with radio observations can answer the questions more clearly when, where and how jets are formed in BHXBs from the observational view. 
   


\begin{ack}
We appreciate Ms.~Catherine Brocksopp that she kindly provided us the radio data used in her excellent paper (\cite{radio}).   The Green Bank Interferometer is a facility of the National Science Foundation operated by the NRAO in support of NASA High Energy Astrophysics programs. We also appreciate an anonymous referee for helpful comments.

TK is supported by the Sumitomo Foundation (2200605).
This research has made use of data and/or software provided by the High Energy Astrophysics Science Archive Research Center (HEASARC),
which is a service of the Astrophysics Science Division at NASA/GSFC. 
\end{ack}


\end{document}